\documentclass[pra,twocolumn,superscriptaddress]{revtex4-1}
\usepackage{float,graphicx,multirow, amsmath,color,dsfont}
\usepackage{amsmath,bm}
\usepackage{amssymb,mathrsfs,dsfont}
\usepackage{amssymb,mathbbol}
\usepackage{amsfonts}
\usepackage{mathrsfs}
\usepackage{float}
\usepackage{xcolor,hyperref}
\definecolor{darkblue}{rgb}{0,0.0.1,0.3}
\definecolor{darkred}{rgb}{0.6,0.1,0}
\hypersetup{colorlinks,breaklinks,
	linkcolor=darkred,urlcolor=darkblue,
	anchorcolor=darkred,citecolor=
	darkred,pdfauthor=ChRiSh, pdftitle=ChRiSh.Estimation}
\usepackage{tikz}
\usepackage{mathbbol}
\usepackage{float}
\usepackage[normalem]{ulem}

\begin{document}
	
	\title{Realistic non-Gaussian operations scheme in parity detection-based Mach-Zehnder quantum interferometry}
\author{Chandan Kumar}
\email{chandan.quantum@gmail.com}
\affiliation{Department of Physical Sciences,
	Indian
	Institute of Science Education and
	Research Mohali, Sector 81 SAS Nagar,
	Punjab 140306 India.}
\author{Rishabh}
\email{rishabh1@ucalgary.ca}
\affiliation{Department of Physics and
	Astronomy, University of Calgary, Calgary T2N1N4, Alberta,
	Canada.}
\author{Shikhar Arora}
\email{shikhar.quantum@gmail.com}
\affiliation{Department of Physical Sciences,
	Indian
	Institute of Science Education and
	Research Mohali, Sector 81 SAS Nagar,
	Punjab 140306 India.}	

\begin{abstract}
	
We theoretically analyze phase sensitivity using parity detection based Mach–Zehnder interferometer (MZI) with the input states generated by  performing non-Gaussian operations, viz., photon subtraction, photon addition, and photon catalysis on a two-mode squeezed vacuum (TMSV) state. Since these non-Gaussian operations are probabilistic, it is of utmost importance to take the success probability into account. To this end, we consider the realistic model of photon subtraction, addition, and catalysis and derive a single expression of the Wigner function for photon subtracted, added, and catalyzed TMSV state. The Wigner function is used to evaluate the lower bound on the phase sensitivity via quantum Cramer-Rao bound and parity detection based phase sensitivity in MZI. We identify the ranges of squeezing and transmissivity parameters where the non-Gaussian states provide better phase sensitivity than the TMSV state. Taking the success probability into account, it turns out that the photon addition is the most advantageous   among all three non-Gaussian operations.
We hope that the generalized Wigner function derived in this work will be useful in various quantum information protocols and state characterization.
	
\end{abstract}
\maketitle

\section{Introduction}

Quantum metrology strives to enhance measurement sensitivity by using quantum resources~\cite{Dowling-cp-2008, Giovannetti2011}.  The canonical example of quantum metrology aims at improving the sensitivity of phase estimation by using a non-classical field of light as input to the Mach-Zehnder interferometer (MZI).  The phase sensitivity of the MZI can reach only up to the short-noise limit (SNL) with solely a classical field of light~\cite{caves-prd-1981}.
On the other hand, with single-mode non-classical states~\cite{Jeong-prl-2019} and entangled states~\cite{Hofmann-pra-2007, Anisimov-prl-2010} as input to MZI, the phase sensitivity can go beyond SNL and reach the Heisenberg limit (HL)~\cite{Giovannetti-science-2004}.  
HL has been reached in photon number parity measurement-based quantum interferometry~\cite{gerry-pra-2000,gerry-pra-2001,gerry-pra-2002,gerry-pra-2003,gerry-pra-2010,gerry-cp-2010,Anisimov-prl-2010,ecs-prl-2011,Seshadreesan_2011,Plick_2010,aravind-2011,sesha-pra-2013,sahota-pra-2013,zhang-pra-2013}, for instance, using N00N states as input~\cite{Hofmann-pra-2007,Dowling-pra-2008,Dowling-cp-2008}. However, the fragility of N00N states in the presence of photon loss limits their utility~\cite{ecs-prl-2011}. Phase sensitivity using a two-mode squeezed vacuum state (TMSV) can even exceed the HL~\cite{Anisimov-prl-2010}. However,  the current experimental techniques pose a challenge to generate a strongly entangled TMSV state~\cite{15dB}.

It has been shown that non-Gaussian operations such as photon subtraction, addition, and catalysis on the TMSV state can enhance the non-classicality and entanglement content of the original state. These non-Gaussian states have been used in various protocols such as quantum teleportation~\cite{tel2000,tel2009,catalysis15,catalysis17,wang2015}, quantum  key distribution~\cite{qkd-pra-2013,qkd-pra-2018,qkd-pra-2019,qk2019,chandan-pra-2019,zubairy-pra-2020},   quantum illumination~\cite{ill2008,ill2013}, and noise-less amplification~\cite{nla-pra-2018} to enhance the performance. With a similar vision, non-Gaussian states have also been considered as input to the MZI to further enhance the phase sensitivity~\cite{gerryc-pra-2012,josab-2012,braun-pra-2014,josab-2016,pra-catalysis-2021}.

These non-Gaussian operations are probabilistic, and therefore it is necessary to consider their success probability. However, the probabilistic nature of these operations has not been taken into account while studying the sensitivity of phase estimation, which can have a significant impact on resource utilization.  This work takes the success probability into account while analyzing the phase sensitivity and identifying the advantageous squeezing parameter and transmissivity region.
To this end, we consider the practical model of photon subtraction, addition, and catalysis~\cite{njp-2015}  and derive
the generalized Wigner function describing the non-Gaussian two mode squeezed vacuum (NG-TMSV) states.  The NG-TMSV states include photon subtracted (PS), photon added (PA), and photon catalyzed (PC) TMSV state.  From hereafter, we use the term non-Gaussian operations (or states) to refer to these three particular non-Gaussian operations (or states) until and unless specified otherwise.  We use the generalized Wigner function of the NG-TMSV state to calculate the quantum Fisher information (QFI) and phase sensitivity of the parity detection based MZI.
We stress that, compared to Gaussian states, the
investigation of these non-Gaussian states  
involves complicated calculations.  Further, the realistic scheme adds an extra complication of transmissivity parameters corresponding to the beam splitters used in the implementation of these non-Gaussian operations,  which significantly enhances the challenge for the theoretical analysis~\cite{njp-2015}.

We analyze the theoretical lower bound on the phase sensitivity for the input NG-TMSV states using Quantum Cramer-Rao bound (QCRB).  We then derive and study the phase sensitivity behavior for these states for parity detection-based MZI.  In order to compare the relative performance of the NG-TMSV states and the TMSV state, we introduce a figure of merit defined as the difference between the phase sensitivity of these states.  This figure of merit enables us to identify the advantageous squeezing and transmissivity parameter ranges.  We also study the impact of the probabilistic nature of the non-Gaussian state generation on the phase sensitivity.   Of the three non-Gaussian operations, the photon addition operation maximizes the product of probability and the difference between the phase sensitivity of the NG-TMSV states and the TMSV state.

The derived Wigner functions for NG-TMSV states, including PSTMSV, PATMSV,  and PCTMSV states, will provide an impetus for dealing with various non-Gaussian CV QIP protocols that generally involve very complex analysis.  Such expressions do not exist in the literature to the best of our knowledge.  Our work also furnishes a single expression of parity detection-based phase sensitivity to cover all three non-Gaussian operations, including symmetric and asymmetric operations.  States generated by ideal symmetric PS, ideal symmetric PA, and asymmetric PC operations on TMSV states have been considered as input to parity detection-based MZI~\cite{josab-2012, josab-2016, pra-catalysis-2021} and form a special case of our general analysis.  The figure of merit defined in this work, along with the considerations involving the probability of non-Gaussian state generation, will allow experimentalists to choose suitable parameters to achieve higher phase sensitivity with resource optimization.

The paper is structured as follows. In Sec.~\ref{cvsystem}, we briefly describe
the formalism of continuous variable systems. In Sec.~\ref{sec:wig}, we derive a general expression of the Wigner  function of the NG-TMSV state. Sec.~\ref{results} contains the analysis of  the lower bound of the phase sensitivity using QCRB. We then study the phase sensitivity using parity detection-based MZI.  Finally, in Sec.~\ref{sec:conc}, we summarize our main results and discuss future prospects.


\section{Formalism of CV systems}
\label{cvsystem}

An $n$-mode
quantum system is represented by  $n$
pairs of Hermitian operator $\hat{q}_i,
\hat{p}_i$ ($i=1,\dots, n$)  known as quadrature
operators~\cite{arvind1995,Braunstein,adesso-2007,weedbrook-rmp-2012,adesso-2014},
which can be written in a column
vector form as 
\begin{equation}\label{eq:columreal}
	\hat{ \xi} =(\hat{ \xi}_i)= (\hat{q}_{1},\,
	\hat{p}_{1} \dots, \hat{q}_{n}, 
	\, \hat{p}_{n})^{T}, \quad i = 1,2, \dots ,2n.
\end{equation}
The canonical commutation relations can be
compactly written  as ($\hbar$=1)
\begin{equation}\label{eq:ccr}
	[\hat{\xi}_i, \hat{\xi}_j] = i \Omega_{ij}, \quad (i,j=1,2,...,2n),
\end{equation}
where $\Omega$ is the 2$n$ $\times$ 2$n$ matrix given by
\begin{equation}
	\Omega = \bigoplus_{k=1}^{n}\omega =  \begin{pmatrix}
		\omega & & \\
		& \ddots& \\
		& & \omega
	\end{pmatrix}, \quad \omega = \begin{pmatrix}
		0& 1\\
		-1&0 
	\end{pmatrix}.
\end{equation}
The quadrature operators are related to the annihilation and creation operators via the relation:
\begin{equation}
	\label{realtocom}
	\hat{a}_i=   \frac{1}{\sqrt{2}}(\hat{q}_i+i\hat{p}_i),
	\quad  \hat{a}^{\dagger}_i= \frac{1}{\sqrt{2}}(\hat{q}_i-i\hat{p}_i).
\end{equation}
It is convenient to describe the CV system in phase space formalism.  
The Wigner distribution for a quantum system with a density
operator $\hat{\rho}$ is defined as
\begin{equation}\label{eq:wigreal}
	W(\bm{\xi}) = \int \frac{\mathrm{d}^n \bm{q'}}{{(2 \pi)}^{n}}\, \left\langle
	\bm{q}-\frac{1}{2}
	\bm{q}^{\prime}\right| \hat{\rho} \left|\bm{q}+\frac{1}{2}\bm{\bm{q}^{\prime}}
	\right\rangle \exp(i \bm{q^{\prime T}}\cdot \bm{p}),
\end{equation}
where
$\bm{\xi} = (q_{1}, p_{1},\dots, q_{n},p_{n})^{T} \in \mathbb{R}^{2n}$,
$\bm{q^{\prime}} \in \mathbb{R}^{n}$ 
and $\bm{q} = (q_1,
q_2, \dots, q_n)^T$, 
$\bm{p} = (p_1, p_2, \dots, p_n)^T $.
The   Wigner function can also be expressed as the
average of displaced parity operator~\cite{parity-1977}:
\begin{equation}\label{wigparity}
	W(\bm{\xi}) =\frac{1}{{ \pi}^{n}} \text{Tr} \left[ \hat{\rho}\, D(\bm{\xi}) \hat{\Pi} D^{\dagger} (\bm{\xi}) \right] ,
\end{equation}
where  $ \hat{\Pi} =\prod_{i=0}^{n}  \exp\left( i \pi   \hat{a}^{\dagger}_i \hat{a}_i \right)$
is the parity operator and $D(\bm{\xi}) = \exp[i \hat{ \xi} \, \Omega \, \bm{\xi}]$ is the displacement operator.
The first-order moments for an $n$ mode system are defined as
\begin{equation}
	\bm{d} = \langle  \hat{\xi}  \rangle =
	\text{Tr}[\hat{\rho} \hat{\xi}],
\end{equation}
and the second-order moments can be written in the form of a  
real symmetric $2n\times2n$ covariance matrix defined as
\begin{equation}\label{eq:cov}
	V = (V_{ij})=\frac{1}{2}\langle \{\Delta \hat{\xi}_i,\Delta
	\hat{\xi}_j\} \rangle,
\end{equation}
where $\Delta \hat{\xi}_i = \hat{\xi}_i-\langle \hat{\xi}_i
\rangle$, and $\{\,, \, \}$ denotes anti-commutator.

A state with a Gaussian Wigner distribution is called a Gaussian state.
For Gaussian states, Wigner function~(\ref{eq:wigreal}) 
can be simplified to~\cite{weedbrook-rmp-2012}
\begin{equation}\label{eq:wignercovariance}
	W(\bm{\xi}) = \frac{\exp[-(1/2)(\bm{\xi}-\bm{d})^TV^{-1}
		(\bm{\xi}-\bm{d})]}{(2 \pi)^n \sqrt{\text{det}V}},
\end{equation}
where  $\bm{d}$ is the displacement and  $V$ denotes the covariance matrix  of the Gaussian state.

Homogeneous symplectic transformations are linear transformations that preserve the canonical commutation relation~(\ref{eq:ccr}). Phase change operation, single-mode squeezing operation, two-mode beam splitter operation, and two-mode squeezing operation are examples of symplectic transformations.	For every homogeneous symplectic transformation $S$, there exists a  corresponding infinite-dimensional unitary representation $\mathcal{U}(S)$ acting on the Hilbert space.
Under such transformations, the density operator transforms as 
$\rho \rightarrow \,\mathcal{U}(S) \rho
\,\mathcal{U}(S)^{\dagger}$.
The corresponding transformation of the displacement vector $\bm{d}$,
covariance matrix $V$ and  Wigner  function is given by~\cite{arvind1995}
\begin{equation}\label{transformation} 
	\bm{d}\rightarrow S \bm{d},\quad V\rightarrow SVS^T,\quad  \text{and} \,\, W(\xi) \rightarrow W(S^{-1} \xi).
\end{equation}

In this work, we will consider non-Gaussian operations on TMSV states.	
A TMSV state is produced by the action of a two-mode squeezing transformation on two uncorrelated vacuum modes.
It is a zero-centered state with the covariance matrix given by	
\begin{equation}
	V_{A_1A_2}=S_{A_1A_2}(r) \mathbb{1}_4 S_{A_1A_2}(r)^T, 
\end{equation}
where $\mathbb{1}_4$ is the $4 \times 4$ identity matrix representing  the covariance matrix of the two uncorrelated
vacuum modes and $S_{A_1A_2}(r)$ is the two-mode squeezing transformation given by
\begin{equation}\label{tms}
	S_{A_1A_2}(r) = \begin{pmatrix}
		\cosh r \,\mathbb{1}_2& \sinh r \,\mathbb{Z} \\
		\sinh r \,\mathbb{Z}& \cosh r \,\mathbb{1}_2
	\end{pmatrix}
	,\quad \mathbb{Z} =  \begin{pmatrix}
		1& 0 \\
		0& -1
	\end{pmatrix},
\end{equation}
where $r$ is the squeezing parameter. The  Wigner function for the TMSV state can be readily computed using Eq.~(\ref{eq:wignercovariance}):
\begin{equation}
	\begin{aligned}
		W(\xi) =  \frac{1}{   \pi  ^2} \exp\big[&- 
		(q_1^2+p_1^2+q_2^2+p_2^2)\cosh (2r) \\
		&+2  (q_1 q_2-p_1 p_2) \sinh (2r)	\big].
	\end{aligned}
\end{equation}

We shall now consider different non-Gaussian operations, viz., photon subtraction, addition, and catalysis, modeled via beam splitters on the  TMSV state.


\section{Wigner  function of non-Gaussian two mode squeezed vacuum  state}\label{sec:wig}

\begin{figure}[H]
	\includegraphics[scale=1]{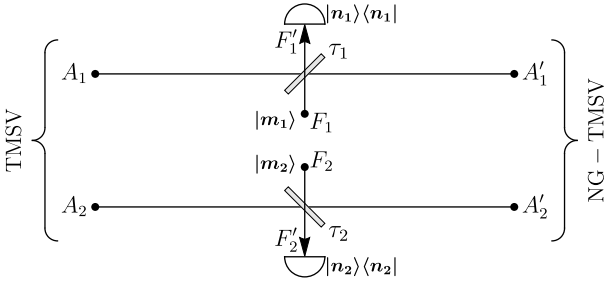}
	\caption{Preparation scheme of non-Gaussian TMSV state.  The TMSV state
		is interfered  with Fock states using beam splitters. Photon number resolving detectors   given by the POVM $\{|n_1\rangle\langle n_1|,\mathbb{1}-|n_1\rangle\langle n_1|\}$
		and $\{|n_2\rangle\langle n_2|,\mathbb{1}-|n_2\rangle\langle n_2|\}$ 
		are applied to modes $F_1^{'}$ and $F_2^{'}$ respectively. }
	\label{figsub}
\end{figure}

The preparation scheme for the NG-TMSV state is shown in Fig.~\ref{figsub}. 
We interfere modes $A_1$ and $A_2$ of the TMSV state  with ancillary modes $F_1$ and $F_2$, initiated to Fock states $|m_1\rangle$ and $|m_2\rangle$, using beam-splitters of transmissivity $\tau_1$ and $\tau_2$ respectively. 
We represent the modes $A_1$ and $A_2$ by the quadrature operators $(\hat{q}_1,\hat{p}_1)^T$ and  $(\hat{q}_2,\hat{p}_2)^T$ and the  auxiliary modes $F_1$ and $F_2$
by the quadrature operators $(\hat{q}_3,\hat{p}_3)^T$ and  $(\hat{q}_4,\hat{p}_4)^T$.
The Wigner function for the four-mode system prior to the beam splitter transformations is given by
\begin{equation}
	W_{F_1 A_1 A_2 F_2}(\xi) =  W_{A_1 A_2}(\xi_1,\xi_2) W_{|m_1\rangle}(\xi_3)  W_{|m_2\rangle}(\xi_4),
\end{equation}
where $\xi_i=(q_i,p_i)^T$ $(i=1,2,3,4)$.
We can evaluate the Wigner  function of a Fock state $|n\rangle$   using Eq.~(\ref{eq:wigreal})   as
\begin{equation}\label{wig:fock}
	W_{|n\rangle}(q,p)=\frac{(-1)^n}{\pi}\exp  \left( 
	-q^2-p^2 \right)\,L_{n}\left[ 2(q^2+p^2) \right].
\end{equation}
The two beam-splitters  $ B(\tau_1,\tau_2)=B_{A_1 F_1}(\tau_1) \oplus B_{A_2 F_2} (\tau_2)$ act on the phase space variables $(\xi_1, \xi_3,\xi_2,\xi_4 )^T$, where the beam-splitter operation $B_{A_k F_k}(\tau_k)$ acting on modes $A_k F_k$   is given by
\begin{equation}\label{beamsplitter}
	B_{A_k F_k}(\tau_k) = \begin{pmatrix}
		\sqrt{\tau_k} \,\mathbb{1}& \sqrt{1-\tau_k} \,\mathbb{1} \\
		-\sqrt{1-\tau_k} \,\mathbb{1}&\sqrt{\tau_k} \,\mathbb{1}
	\end{pmatrix},\quad (k=1,2).
\end{equation}
The transformed Wigner function is given by
\begin{equation}
	\begin{aligned}
		W_{F_1' A_1' A_2' F_2'}(\xi)   =W_{F_1 A_1 A_2 F_2}( B(\tau_1,\tau_2)^{-1}\xi)  .
	\end{aligned}
\end{equation}
The modes $F_1^{'}$ and $F_2^{'}$ are measured using
photon number resolving detectors (PNRD), given by the positive-operator-valued measure (POVM) $\{\Pi_{n_1}=|n_1\rangle\langle n_1|,\mathbb{1}-\Pi_{n_1}\}$ and $\{\Pi_{n_2}=|n_2\rangle\langle n_2|,\mathbb{1}-\Pi_{n_2}\}$, respectively. The simultaneous click of the POVM elements $\Pi_{n_1}$ and $\Pi_{n_2}$ heralds successful non-Gaussian operations  on both the modes. 
The corresponding  unnormalised Wigner   function is given by
\begin{equation}\label{detect}
	\begin{aligned}
		\widetilde{W}^{\text{NG}}_{A_1' A_2'}(\xi_1,\xi_2)=& (2 \pi)^2\int  d^2 \xi_3 d^2 \xi_4  \underbrace{W_{F_1' A_1' A_2' F_2'}(\xi_1,\xi_2,\xi_3,\xi_4)}_{\text{Four mode entangled state}}\\
		&\times 
		\underbrace{W_{|n_1\rangle }(\xi_3)}_{\text{Projection on }
			|n_1\rangle \langle n_1|} 
		\underbrace{W_{|n_2\rangle }(\xi_4 )}_{\text{Projection on }|n_2\rangle \langle n_2|}. \\
	\end{aligned}
\end{equation}

By choosing suitable values of $(m_i, n_i)$, we can perform three different
non-Gaussian operations on mode $A_i$ as following: (i) photon subtraction for	  $m_i < n_i$,
(ii) photon addition for   $m_i > n_i$, and (iii) photon catalysis for   $m_i = n_i$. 

The action of  photon subtraction, photon addition, and photon catalysis on TMSV states yields PSTMSV, PATMSV, and PCTMSV states, respectively, which are non-Gaussian states. However, zero-photon catalysis, corresponding to   $m_i = n_i =0$, is a Gaussian operation, and therefore, the resulting state, zero-photon catalyzed TMSV state,  is a Gaussian state.

In this work, we consider both asymmetric and symmetric non-Gaussian operations on TMSV state, which can be obtained by putting suitable   conditions on parameters $m_i$, $n_i$, and $\tau_i$, as shown in Table~\ref{table1}.
It should be noted that the asymmetric non-Gaussian operations are performed on mode $A_2$ of the TMSV state. 

\begin{table}[h!]
	\centering
	\caption{\label{table1}
		Conditions on the number of input photons $m_i$, detected photons $n_i$ and the transmissivity $\tau_i$ of the beam splitters for various asymmetric and symmetric non-Gaussian operations on the TMSV state.}
	\setlength{\tabcolsep}{7pt}
	\renewcommand{\arraystretch}{1}	
	\begin{tabular}{ |c|c c|c c|c c|}
		\hline 
		\multirow{2}{*}{Operations} & \multicolumn{2}{c|}{Input} &  \multicolumn{2}{c|}{Detected} & \multicolumn{2}{c|}{Transmissivity}\\
		\cline{2-7}
		& $m_1$ & $m_2$ & $n_1$ & $n_2$ & \quad $\tau_1$ &  $\tau_2$  \\
		\hline \hline
		Asym $n$-PS  & 0& 0& 0&$n$ & \multirow{3}{*}{ \quad $1$} & \multirow{3}{*}{$\tau$} \\ \cline{1-5}
		Asym $n$-PA  & 0 & $n$ & 0 & 0 & &\\ \cline{1-5}
		Asym $n$-PC  & 0 & $n$ & 0&$n$ & & \\ \hline  \hline
		Sym $n$-PS 	 & 0& 0& $n$ &$n$ & \multirow{3}{*}{ \quad $\tau$} & \multirow{3}{*}{$\tau$} \\ \cline{1-5}
		Sym $n$-PA 	 & $n$ & $n$ & 0 & 0 & & \\  \cline{1-5}
		Sym $n$-PC 	 & $n$ & $n$ & $n$ & $n$ & & \\  \hline \hline
	\end{tabular}
\end{table}

Equation~(\ref{detect}) can be converted into a Gaussian integral
using the generating function for the Laguerre polynomial appearing in the Wigner function
of the Fock state~(\ref{wig:fock}):
\begin{equation}\label{gen}
	\begin{aligned}
		L_n[2(q^2+p^2)]=	\bm{\widehat{D}}\exp \left[\frac{st}{2}+s(q+ip)-t(q-ip)\right],
	\end{aligned}
\end{equation}
with
\begin{equation}
	\bm{\widehat{D}} = \frac{2^n}{n!}  \frac{\partial^n}{\partial\,s^n} \frac{\partial^n}{\partial\,t^n} \{ \bullet \}_{s=t=0}.
\end{equation}
Integration of Eq.~(\ref{detect}) yields 
\begin{equation}\label{eq4}
	\widetilde{W}^{\text{NG}}_{A_1' A_2'}=  \frac{1}{a_0 \pi^2}   \bm{\widehat{D}_1} \exp \left(\bm{\xi}^T M_1 \bm{\xi}+\bm{u}^T M_2 \bm{\xi} + \bm{u}^T M_3 \bm{u} \right) ,
\end{equation}
where $a_0 =1+\alpha^2(1-\tau_1 \tau_2)  $, column vectors $\bm{\xi}$ and $\bm{u}$ are defined as
\begin{equation}
	\begin{aligned}
		\bm{\xi}=&(q_1,p_1,q_2,p_2)^T, \\
		\bm{u}=&(u_1,v_1,u_2,v_2,u_1',v_1',u_2',v_2')^T, \\
	\end{aligned}
\end{equation}
and differential operator $\bm{\widehat{D}_1} $ is defined as 
\begin{multline}
	\bm{\widehat{D}_1} = \frac{(-2)^{m_1+m_2+n_1+n_2}}{m_1!m_2!n_1!n_2!} \frac{\partial^{m_1}}{\partial\,u_1^{m_1}} \frac{\partial^{m_1}}{\partial\,v_1^{m_1}} \frac{\partial^{m_2}}{\partial\,u_2^{m_2}} \frac{\partial^{m_2}}{\partial\,v_2^{m_2}}\\
	\times \frac{\partial^{n_1}}{\partial\,u_1'^{n_1}} \frac{\partial^{n_1}}{\partial\,v_1'^{n_1}} \frac{\partial^{n_2}}{\partial\,u_2'^{n_2}} \frac{\partial^{n_2}}{\partial\,v_2'^{n_2}} \{ \bullet \}_{\substack{u_1= v_1=u_2= v_2=0\\ u_1'= v_1'=u_2'= v_2'=0}}.\\
\end{multline}
Further, the explicit form of the matrices $M_1$, $M_2$, and $M_3$ are provided in Eqs.~(\ref{mat1}), (\ref{mat2}), and~(\ref{mat3}) of Appendix~\ref{appwigner}.
The probability  of   $n_1$ and $n_2$ photon detection on mode $F'_1$ and $F'_2$, respectively, can be evaluated as
\begin{equation}\label{successp}
	\begin{aligned}
		P^{\text{NG}}= & \int d^2 \xi_1 d^2 \xi_2 \widetilde{W}^{\text{NG}}_{A_1' A_2'}(\xi_1,\xi_2),\\
		= & a_0^{-1}\bm{\widehat{D}_1} \exp \left(\bm{u}^T M_4 \bm{u}\right),\\
	\end{aligned}
\end{equation}
where the matrix $M_4$  is given in  Eq.~(\ref{mat4}) of Appendix~\ref{appprob}.
\begin{figure*}
	\includegraphics[scale=1]{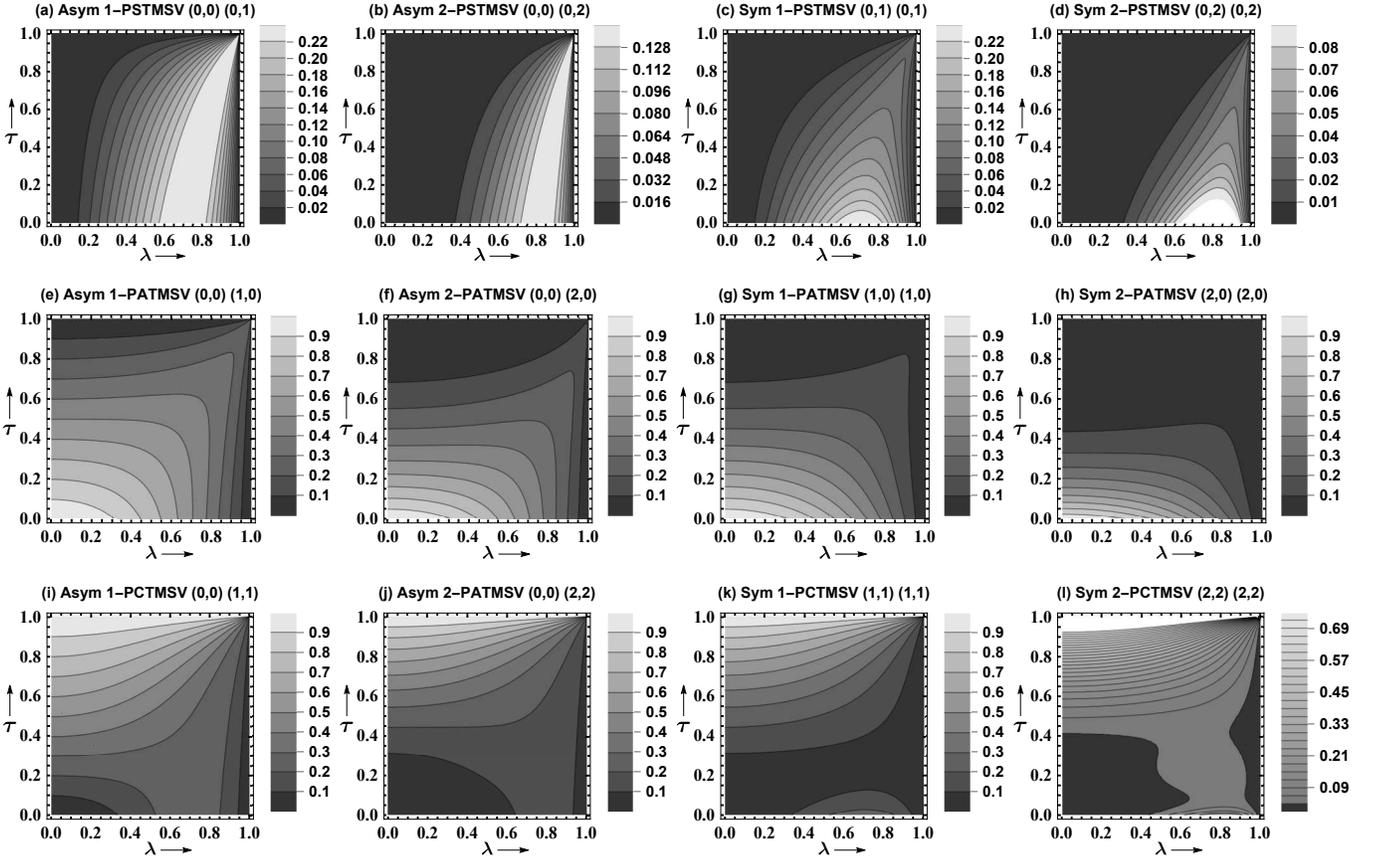}
	\caption{The probability $P^{\text{NG}}$ of detecting $n_1$ and $n_2$ photons on modes $A_1$ and $A_2$ as a   function of the transmissivity $\tau$ and
		squeezing parameter $\lambda$. For symmetric non-Gaussian operations, we have set $\tau_1=\tau_2=\tau$.  For asymmetric  non-Gaussian operations, $\tau_1 = 1$ and    $\tau_2=\tau$. The value of $(m_1,n_1)(m_2,n_2)$ for each panel is also shown.}
	\label{ngprob}
\end{figure*}
Figure~\ref{ngprob} shows the success probability of various non-Gaussian operations, which is the same as the probability of detecting $n_1$ and $n_2$ photons on modes $A_1$ and $A_2$, as a function of the transmissivity $\tau$ and squeezing parameter $\lambda = \tanh \,r$.  
We notice a vertical band of high success probability for asymmetric photon subtraction for intermediate squeezing values and all transmissivity values. In contrast, for symmetric photon subtraction, the region of high success probability occurs only for low transmissivity and intermediate values of squeezing.  
On the other hand, for photon addition, the success probability is high for low transmissivity and small squeezing. On the contrary, we notice a horizontal band of high success probability for high transmissivity values and all squeezing values for photon catalysis. 
Interestingly, in the limit $\tau \rightarrow 1$, the incoming photon is detected with unit probability, and therefore, the success probability for photon catalysis approaches unity.
However, the resulting state is the same as the input TMSV state, and therefore, no catalysis operation takes place.

Of all the three non-Gaussian operations, photon subtraction occurs with relatively low probability compared to photon addition and catalysis. We notice two general trends: (i) success probability of non-Gaussian operations on both the modes is less as compared to non-Gaussian operations on one of the modes; (ii) success probability decreases for higher photon number detection.

The normalized  Wigner  function $W^{\text{NG}}_{A'_1 A'_2}$
of the NG-TMSV state turns out to be
\begin{equation}\label{normPS}
	W^{\text{NG}}_{A'_1 A'_2}(\xi_1,\xi_2) ={\left(P^{\text{NG}}\right)}^{-1}\widetilde{W}^{\text{NG}}_{A_1' A_2'}(\xi_1,\xi_2).
\end{equation}

We can easily obtain several special cases from the aforederived Wigner function of the NG-TMSV state.
For instance, the Wigner function of the ideal PSTMSV state $\hat{a}_1^{n_1} \hat{a}_2^{n_2}  |\text{TMSV}\rangle$ can be obtained by setting $\tau_1=\tau_2=1$ in the symmetric photon subtraction case. Similarly, the Wigner function of the ideal PATMSV state 
$\hat{a}{_1^{\dagger }}^{m_1}\hat{a}{_2^{\dagger }}^{m_2}  |\text{TMSV}\rangle$
can be obtained by setting $\tau_1=\tau_2=1$ in the symmetric photon addition case.

We can calculate the average of  Weyl (symmetrically) ordered operators using the Wigner function as follows:
\begin{equation}
	\left\langle  {}_{\bm{:}}^{\bm{:}} \hat{q_1}^{a_1} \hat{p_1}^{b_1} \hat{q_2}^{a_2} \hat{p_2}^{b_2} {}_{\bm{:}}^{\bm{:}} \right\rangle = \int d^4 \xi\, q_1^{a_1} p_1^{b_1} q_2^{a_2} p_2^{b_2} W^{\text{NG}}_{A'_1 A'_2}(\xi ),
\end{equation}
where the symbol ${}_{\bm{:}}^{\bm{:}}  \bullet {}_{\bm{:}}^{\bm{:}} $  represents Weyl ordering.
This quantity, akin to moment generating function, can be evaluated using
parametric differentiation technique as follows:
\begin{equation}\label{para}
	\mathcal{M}_{a_1,b_1}^{a_2,b_2}= \bm{\widehat{D}_2} \int d^4 \xi\,
	e^{  x_1 q_1+y_1 p_1+x_2 q_2+y_2p_2 }
	W^{\text{NG}}_{A'_1 A'_2}(\xi ),
\end{equation}
with
\begin{equation}
	\bm{\widehat{D}_2} =   \frac{\partial^{a_1}}{\partial\, x_1^{a_1}} \frac{\partial^{b_1}}{\partial\, y_2^{b_2}}\frac{\partial^{a_2}}{\partial\, x_2^{a_2}} \frac{\partial^{b_2}}{\partial\, y_2^{b_2}} \{ \bullet \}_{x_1=y_1=x_2=y_2=0}.
\end{equation}
On integrating Eq.~(\ref{para}), we obtain
\begin{equation}\label{momgen}
	\mathcal{M}_{a_1,b_1}^{a_2,b_2}	=\frac{\bm{\widehat{D}_2 \widehat{D}_1} \exp \left(\bm{u}^T M_4 \bm{u}+\bm{u}^T M_5 \bm{x} + \bm{x}^T M_6 \bm{x}  \right)}{\bm{\widehat{D}_1} \exp \left(\bm{u}^T M_4 \bm{u}\right)},
\end{equation}
where  $\bm{x}=(x_1,y_1,x_2,y_2)^T$ is a column vector, and the explicit form of matrices $M_5$  and $M_6$ are provided in Eqs.~(\ref{mat5})   and~(\ref{mat6}) of Appendix~\ref{appmomgen}.


\section{Phase estimation with the NG-TMSV state via MZI   }\label{results}
\begin{figure}[H]
	\begin{center}
		\includegraphics[scale=1]{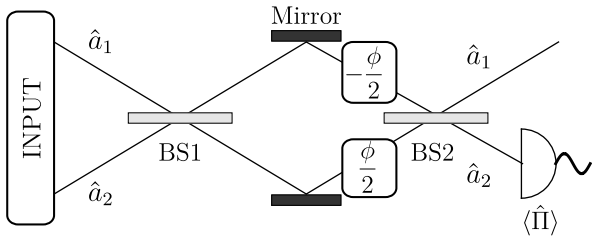}
		\caption{ Schematic of the Mach-Zehnder interferometer for the phase shift detection.}
		\label{mzi}
	\end{center}
\end{figure}
We consider a balanced MZI consisting of two $50:50$ beam splitters and two-phase shifters, as depicted in Fig.~\ref{mzi}. The two input modes are denoted by the annihilation operators $\hat{a}_1$ and $\hat{a}_2$. The input to the interferometer is NG-TMSV states, including PSTMSV, PATMSV, and PCTMSV states. 
 It should be noted that the cases corresponding to unsuccessful non-Gaussian operations are discarded.
Here we use the well-known Schwinger representation of $\text{SU}(2)$ algebra to describe the transformation of a beam splitter~\cite{yurke-1986}. The generators of the $\text{SU}(2)$ algebra can be described using the two sets of Bose operators as
\begin{equation}
	\begin{aligned}
		\hat{J}_1 = &\frac{1}{2}(\hat{a}^\dagger_1\hat{a}_2+\hat{a}_1\hat{a}^\dagger_2),\\
		\hat{J}_2 = &\frac{1}{2i}(\hat{a}^\dagger_1\hat{a}_2-\hat{a}_1\hat{a}^\dagger_2),\\
		\hat{J}_3 = &\frac{1}{2}(\hat{a}^\dagger_1\hat{a}_1-\hat{a}^\dagger_2\hat{a}_2),
	\end{aligned}
\end{equation}
which satisfy the commutation relations $[J_i,J_j] = i \epsilon_{ijk}J_k $.
While the action of the first and the second balanced  beam splitters are given by $e^{-i(\pi/2)J_1}$ and $e^{i(\pi/2)J_1}$, the collective action of the two phase shifters is given by $e^{i \phi J_3}$. 
Therefore, the infinite-dimensional unitary transformation corresponding to the balanced MZI can be written as
\begin{equation}
	\mathcal{U}(S_{\text{MZI}}) = e^{-i(\pi/2)J_1}e^{i \phi J_3}e^{i(\pi/2)J_1}=e^{-i \phi J_2},
\end{equation}
where $\phi$ is the unknown phase to be estimated.
The corresponding symplectic transformation $S_{\text{MZI}}$ acting on the phase space variables $(\xi_1,\xi_2)^T$ is given by
\begin{equation} 
	S_{\text{MZI}} = \begin{pmatrix}
		\cos (\phi/2) \,\mathbb{1}& -	\sin (\phi/2) \,\mathbb{1} \\
		\sin (\phi/2) \,\mathbb{1}& \cos (\phi/2) \,\mathbb{1}
	\end{pmatrix}.
\end{equation}
Therefore, the input Wigner function transforms as follows under the action of $S_{\text{MZI}}$:
\begin{equation}
	W_{\text{in}}(\xi ) \rightarrow W_{\text{in}}(S_{\text{MZI}}^{-1}\xi) =W_{\text{out} } (\xi).
\end{equation}

\subsection{Quantum Fisher information}

Although we will be using parity detection to estimate the phase,
QCRB provides a useful lower bound of the phase sensitivity.  This lower bound of phase sensitivity is given by~\cite{braunstein-prl-1994}
\begin{equation}
	\Delta \phi _{\text{min}} = \frac{1}{\sqrt{F_Q}},
\end{equation}
where $F_Q$ is QFI. It is independent of the type of measurement performed and depends solely on the input state. It can be calculated for a pure state as follows:
\begin{equation}
	F_Q= 4\left[ \langle \psi^{\prime}|\psi^{\prime} \rangle
	- \langle \psi^{\prime}|\psi  \rangle \right],
\end{equation}
where $|\psi  \rangle = e^{i\phi J_3}e^{i\pi J_1/2}|\text{in}\rangle $
is the quantum state prior to the second beam splitter and $|\psi^{\prime} \rangle = \partial |\psi\rangle/\partial \phi$.
The QFI can also be written in the term of the input state as
\begin{equation}\label{fisher}
	F_Q=	4\left[ \langle\text{in} |\hat{J}_2^2|\text{in} \rangle -|\langle\text{in} |\hat{J}_2|\text{in} \rangle|^2\right].
\end{equation}

To evaluate the QFI using the moment generating function~(\ref{momgen}), we write $\hat{J}_{2}$ and $\hat{J}^{2}_{2}$ in terms of the quadrature operators and symmetrize them. The operator
\begin{equation}
	\hat{J}_{2} = \frac{1}{2}(\hat{q}_1\hat{p}_2-\hat{p}_1\hat{q}_2),
\end{equation}
is already symmetric in the quadrature operators. We note that for the NG-TMSV states, $\langle\text{in} |\hat{J}_2|\text{in} \rangle$ evaluates to zero.
The operator$\hat{J}^{2}_{2}$ can be written as
\begin{equation}
	\hat{J}^{2}_{2} = \frac{1}{4}(\hat{q}^{2}_1\hat{p}^{2}_2+\hat{p}^{2}_1\hat{q}^{2}_2-\hat{q}_1\hat{p}_1\hat{p}_2\hat{q}_2
	-\hat{p}_1\hat{q}_1\hat{q}_2\hat{p}_2).
\end{equation}
On symmetrizing $\hat{J}^{2}_{2}$, we get
\begin{equation}
	\begin{aligned}
		\hat{J}^{2}_{2}&  = \frac{1}{4}\bigg[\hat{q}^{2}_1\hat{p}^{2}_2+\hat{p}^{2}_1\hat{q}^{2}_2\\
		&	-\frac{(\hat{q}_1\hat{p}_1+\hat{p}_1\hat{q}_1+i)}{2}\frac{(\hat{p}_2\hat{q}_2+\hat{q}_2\hat{p}_2-i)}{2}\\
		&	-\frac{(\hat{q}_1\hat{p}_1+\hat{p}_1\hat{q}_1-i)}{2}\frac{(\hat{p}_2\hat{q}_2+\hat{q}_2\hat{p}_2+i)}{2}
		\bigg].
	\end{aligned}
\end{equation}
Therefore, the QFI~(\ref{fisher}) can be written as
\begin{equation}
	\begin{aligned}
		F_Q= -\frac{1}{8}+\frac{1}{4}\langle \hat{q}^{2}_1\hat{p}^{2}_2\rangle
		+\frac{1}{4}\langle \hat{p}^{2}_1\hat{q}^{2}_2\rangle -\frac{1}{2}\langle 	{}_{\bm{:}}^{\bm{:}} \hat{q}_1 \hat{p}_1 \hat{q}_2  \hat{p}_2	{}_{\bm{:}}^{\bm{:}}\rangle.
	\end{aligned}
\end{equation}
This can be easily evaluated using the moment generating function~(\ref{momgen}) as
\begin{equation}
	F_Q= 	 -\frac{1}{8}+\frac{1}{4}\mathcal{M}_{2,0}^{0,2}  
	+\frac{1}{4}\mathcal{M}_{0,2}^{2,0}   -\frac{1}{2}\mathcal{M}_{1,1}^{1,1}  .
\end{equation}
\begin{figure} 
	\begin{center}
		\includegraphics[scale=1]{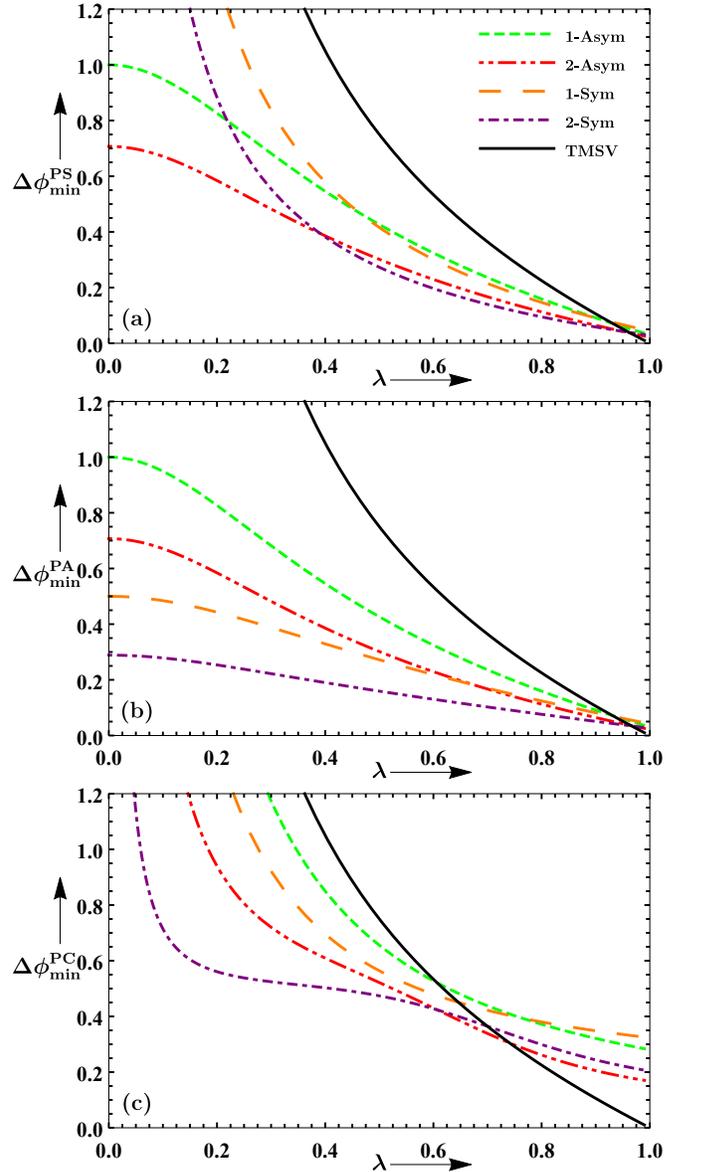}
		\caption{The minimum phase uncertainty $\Delta \phi_{\text{min}}$ obtained from the quantum Cramer-Rao bound, for NG-TMSV states,  as a function of the  squeezing parameter $\lambda$.  The value of transmissivity has been taken as $\tau=0.9$ for (a) and (b) and $\tau=0.2$  for (c). }
		\label{fisher_1d_l}
	\end{center}
\end{figure}
We first analyze  the effect of squeezing on $\Delta \phi _{\text{min}}$,
while the  transmissivity is kept fixed. We plot $\Delta \phi _{\text{min}}$ as a function of squeezing in Fig.~\ref{fisher_1d_l}.

The results show that $\Delta \phi _{\text{min}}$ for NG-TMSV states can achieve a lower value as compared to TMSV state.
Among all the three non-Gaussian operations, symmetric photon addition attains the minimum value of $\Delta \phi _{\text{min}}$.
Since the expressions for $\Delta \phi _{\text{min}}$ corresponding to the Asym $n$-PSTMSV and Asym $n$-PATMSV states are the same, they yield the same results as can be seen in the plots. Within asymmetric operations, $\Delta \phi_{\text{min}}$ achieves a lower value for higher photon number detection, and the same is true for symmetric non-Gaussian operations. However, as noticed in the previous section, the probability decreases for higher photon number detection.

\begin{figure} 
	\begin{center}
		\includegraphics[scale=1]{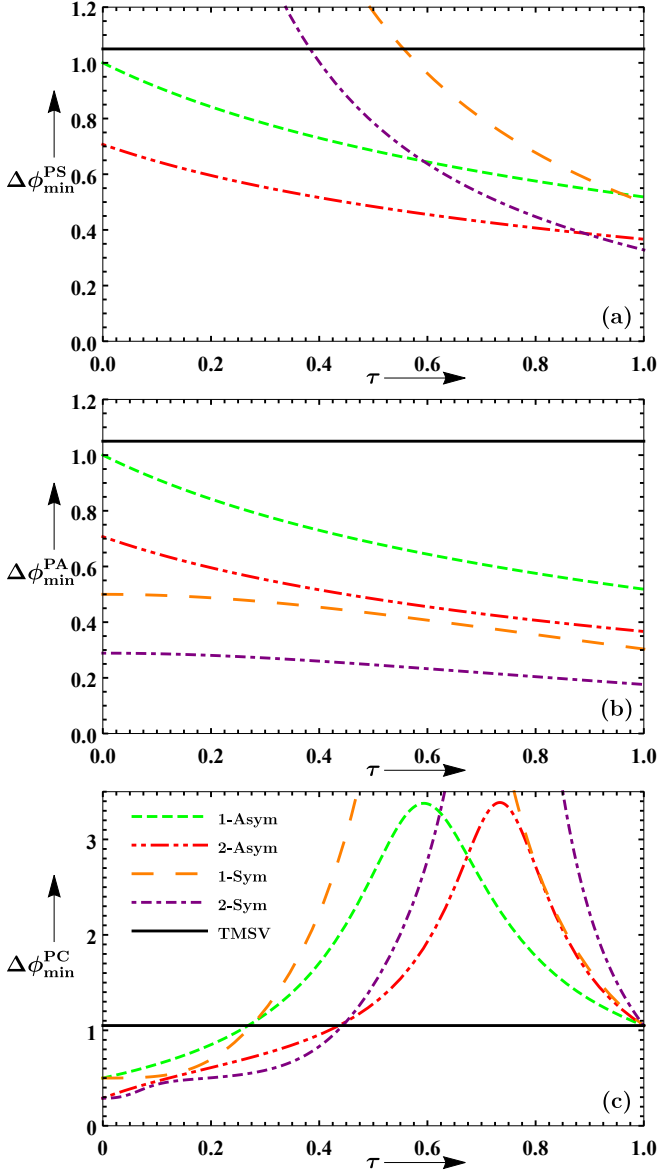}
		\caption{The minimum phase uncertainty $\Delta \phi_{\text{min}}$ obtained from the quantum Cramer-Rao bound, for NG-TMSV states,  as a function of transmissivity $\tau$.  The value of squeezing parameter has been set as $\lambda=0.4$. }
		\label{fisher_1d_t}
	\end{center}
\end{figure}
We now analyze the effect of transmissivity on $\Delta \phi _{\text{min}}$ while keeping the squeezing fixed.   The result is shown in Fig.~\ref{fisher_1d_t}.
For photon subtraction and addition operations, $\Delta\phi_{\text{min}}$ is minimized in the limit $\tau \rightarrow 1$, while for photon catalysis, $\Delta\phi_{\text{min}}$ is minimized in the limit $\tau \rightarrow 0$. However, the probability for photon subtraction and addition approaches zero, in the limit $\tau \rightarrow 1$, and the probability of photon catalysis is low in the limit $\tau \rightarrow 0$. Again we observe that $\Delta \phi_{\text{min}}$ achieves a lower value for higher photon number detection in the case of both asymmetric and symmetric photon addition and subtraction. However, this is only true in the low transmissivity regime for photon catalysis.

\subsection{Parity detection based phase sensitivity}

In this work, we measure the photon number parity operator on the output mode $\hat{a}_2$ to estimate the phase. The corresponding photon number parity operator is given by
\begin{equation}
	\hat{\Pi}_{\hat{a}_2} =  \exp\left( i \pi   \hat{a}^{\dagger}_2 \hat{a}_2 \right)=  (-1)^{\hat{a}_2^{\dagger}\hat{a}_2}.
\end{equation}
This measurement differentiates between odd and even numbers of photons. The expectation value of the parity operator can be written in terms of the Wigner function using Eq.~(\ref{wigparity}) as~\cite{Birrittella-2021}
\begin{equation}
	\langle \hat{\Pi}_{\hat{a}_2} \rangle =f(\phi)  = \pi \int \, d^2\xi_1 \, W_{\text{out}} (\xi_1,0).
\end{equation}
Using the Wigner function of the input NG-TMSV state~(\ref{normPS}), the average of the parity operator evaluates to
\begin{equation}\label{apari}
	f(\phi) =\frac{ a_0 \,  \bm{\widehat{D}_1} \exp \left(\bm{u}^T M_7 \bm{u} \right)}{b_0 \, \bm{\widehat{D}_1} \exp \left(\bm{u}^T M_4 \bm{u}\right)},
\end{equation}
where $b_0 = (1-\lambda ^2)^{-1}\sqrt{1+\lambda ^2 \tau _1 \tau _2 \left(\lambda ^2 \tau _1 \tau _2+2 \cos (2 \phi )\right)}$ 
and the explicit form of matrix $M_7$ is provided in Eq.~(\ref{mat7}) of Appendix~\ref{apppari}.

The phase uncertainty or sensitivity can be obtained using the error propagation formula as
\begin{equation}
	\Delta \phi = \frac{\sqrt{1-f(\phi+\pi/2) ^2}}{|\partial  f(\phi+\pi/2) /\partial \phi|}.
\end{equation}


\begin{figure}
	\begin{center}
		\includegraphics[scale=1]{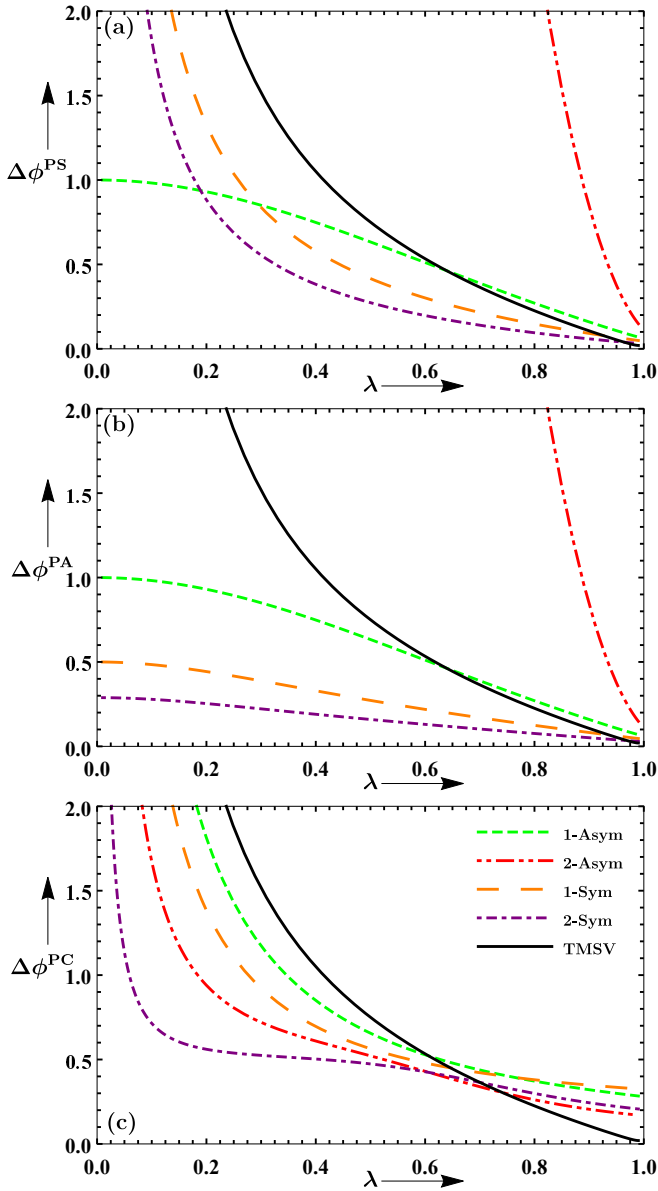}
		\caption{Phase uncertainty $\Delta \phi $   for NG-TMSV states,  as a function of the  squeezing parameter $\lambda$.  The value of transmissivity has been taken as $\tau=0.9$ for (a) and (b) and $\tau=0.2$  for (c), while phase has been set as $\phi=0.01$ for all the cases.}
		\label{phase_1d_l}
	\end{center}
\end{figure}
We now numerically investigate the dependence of $\Delta \phi $ on squeezing, transmissivity, and phase magnitude.
First, we plot $\Delta \phi$ as a function of squeezing while keeping the transmissivity and phase constant. The result is shown in Fig.~\ref{phase_1d_l}. 
Symmetric photon subtraction and addition perform better than TMSV state for almost the whole range of squeezing, but the relative performance compared to TMSV state becomes worse as $\lambda  $ approaches one. 
Among asymmetric cases, single-photon subtraction and addition yield better phase sensitivity only up to a certain threshold squeezing above which the TMSV state performs better. Similar behavior is also observed for all symmetric and asymmetric photon catalysis cases.

We also notice a few similarities between Figs.~\ref{fisher_1d_l} and~\ref{phase_1d_l}:
(i) the qualitative behavior for different non-Gaussian operations are identical except for asymmetric subtraction and addition operations, (ii) the performance of symmetric photon addition is the best among all the non-Gaussian operations, (iii) the relative performance of NG-TMSV states as compared to TMSV state enhances for small values of squeezing, (iv) both $\Delta \phi_{\text{min}}$ and $\Delta \phi$ achieve lower values for higher photon number detection in the case of asymmetric and symmetric non-Gaussian operations except for asymmetric photon subtraction and addition. We note that Asym 2-PSTMSV and Asym 2-PATMSV states never yield phase sensitivity better than the TMSV state.

\begin{figure}
	\begin{center}
		\includegraphics[scale=1]{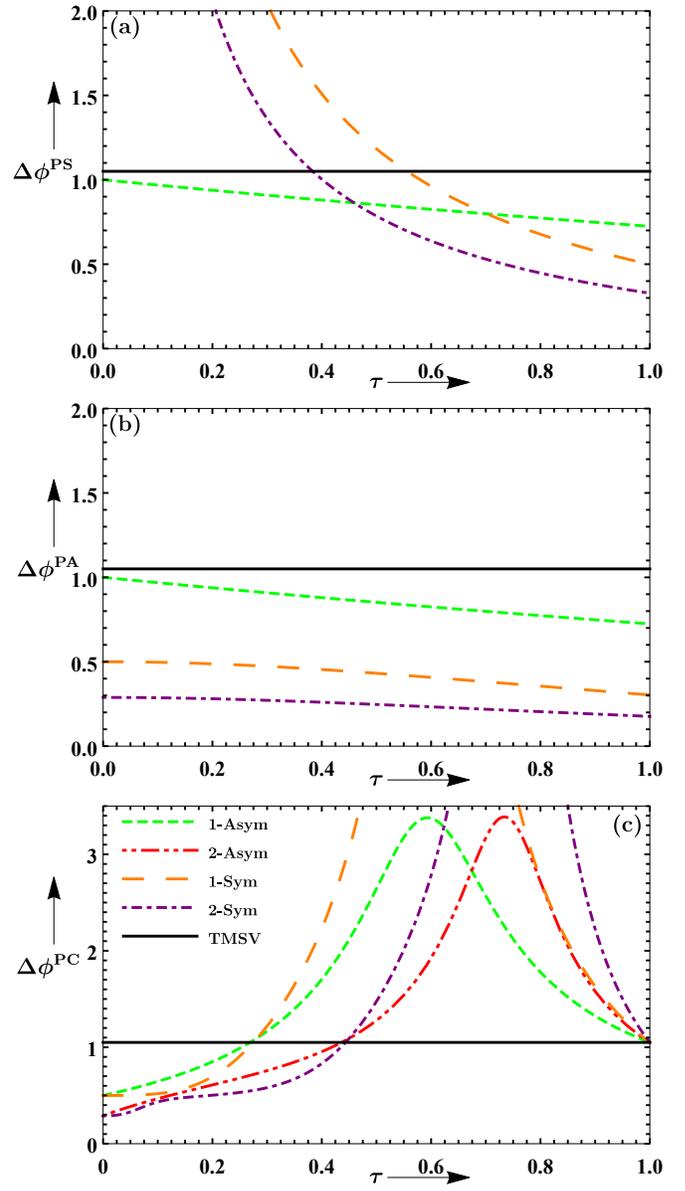}
		\caption{  Phase uncertainty $\Delta \phi $   for NG-TMSV states,  as a function of the  beam splitter transmissivity $\tau$.  The parameters have been set  as $\lambda=0.4$ and $\phi=0.01$ for all the cases.}
		\label{phase_1d_t}
	\end{center}
\end{figure}
We now plot $\Delta \phi$ as a function of transmissivity for fixed squeezing and phase in Fig.~\ref{phase_1d_t}.
As can be seen in Fig.~\ref{fisher_1d_t}, $\Delta\phi$ is minimized in the limit $\tau \rightarrow 1$ for photon subtraction and addition operations, while for photon catalysis, $\Delta\phi$ is minimized in the limit $\tau \rightarrow 0$. The qualitative behavior for different non-Gaussian operations is also similar to Fig.~\ref{fisher_1d_t} except for the cases of Asym  2-PSTMSV and Asym  2-PATMSV states. These two states do not appear in the graph because their phase sensitivities lie far above the plot range.

\begin{figure}
	\begin{center}
		\includegraphics[scale=1]{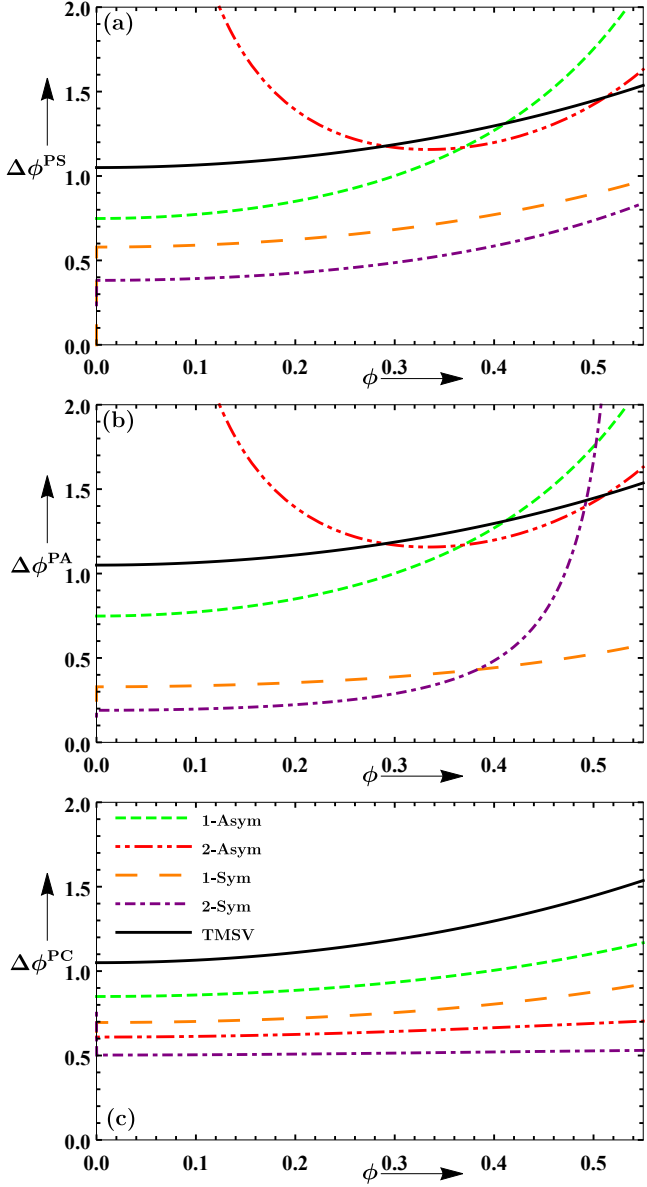}
		\caption{Phase uncertainty $\Delta \phi $   for NG-TMSV states,  as a function of the  phase $\phi$.  The value of transmissivity has been taken as $\tau=0.9$ for (a) and (b) and $\tau=0.2$  for (c), while squeezing parameter has been set as $\lambda=0.4$ for all the cases.}
		\label{phase_1d_p}
	\end{center}
\end{figure}

In Fig.~\ref{phase_1d_p}, we show the plot of $\Delta \phi$ as a function of phase for fixed transmissivity and squeezing. We observe that photon catalysis enhances the phase sensitivity even for larger $\phi$. Cross over between different PSTMSV and PATMSV states happens for larger values of phase, for instance, Sym 2-PATMSV state crosses over Sym 1-PATMSV state at $\phi \approx 0.4$. Furthermore, Asym 2-PSTMSV and Asym 2-PATMSV states perform better than TMSV for a brief interval of $\phi$.

\subsection{Relative performance of NG-TMSV states}
\label{subsec:tf}
We now proceed to study the relative performance of the NG-TMSV states compared to the TMSV state.
To this end, we define
a figure of merit, $\mathcal{D}^{\text{NG} }$, as the difference of $\Delta\phi$ between TMSV and NG-TMSV state:  
\begin{equation}
	\mathcal{D}^{\text{NG} }= \Delta\phi^\text{TMSV}- \Delta\phi^\text{NG-TMSV}.
\end{equation}
This figure of merit enables us to identify the parameter region of transmissivity and squeezing where the NG-TMSV states perform better than the TMSV state. This corresponds to region of a positive $\mathcal{D}^{\text{NG} }$.

 We note that the success probability, which represents the fraction of successful non-Gaussian operations per trial, quantifies the resource utilization. We can encounter scenarios where $\mathcal{D^{\text{NG}}}$  is large; however, the success probability is low  representing a poor resource utilization. Therefore, it is better to maximize  the product $\mathcal{D^{\text{NG}}} \times P^{\text{NG}}$ rather $\mathcal{D^{\text{NG}}}$. We first qualitatively take the probabilistic nature into account and then proceed to a quantitative analysis of the same.

 We now plot $\mathcal{D^{\text{NG}}}$ for various non-Gaussian states as a function of the transmissivity $\tau$ and squeezing parameter $\lambda$.
\begin{figure}[H] 
	\begin{center}
		\includegraphics[scale=1]{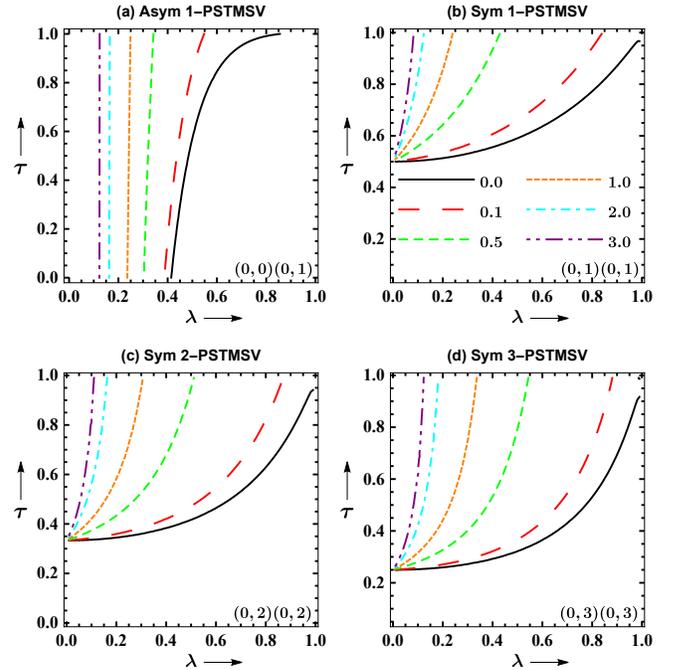}
		\caption{Plots of    fixed $\mathcal{D^{\text{PS}}}$, difference of $\Delta \phi$ between TMSV and PSTMSV states,  as a function of the transmissivity $\tau$ and
			squeezing parameter $\lambda$.   The value of $(m_1,n_1)(m_2,n_2)$ has been shown in the bottom right. We have set the phase $\phi=0.01$ for all the cases.
			The labels in panel (b) correspond to plotted values of $\mathcal{D^{\text{PS}}}$. 	}
		\label{phase_sub_c}
	\end{center}
\end{figure}

Figure~\ref{phase_sub_c} shows the plot of various fixed values of $\mathcal{D^{\text{PS}}}$~$(=0.0,\,0.1,\,0.5,\,1,\,2,\,3)$ as a function of the transmissivity $\tau$ and squeezing parameter $\lambda$. Regions of $(\tau,\lambda)$ with positive values of $\mathcal{D^{\text{PS}}}$ indicate that the PSTMSV states perform better than the TMSV state.  The loci of $\mathcal{D^{\text{PS}}}=0$ progresses along those values of $(\tau, \lambda)$, for which the phase sensitivity of the PSTMSV state is equal to the TMSV state; however, at those specific values of $(\tau, \lambda)$, the PSTMSV state is not the same as TMSV state.
The positive region of $\mathcal{D^{\text{PS}}}$ for Asym 1-PSTMSV state occurs for squeezing below $\lambda \approx 0.4$ for all values of transmissivity. Decreasing the squeezing results in the enhancement of $\mathcal{D^{\text{PS}}}$. However, as we can see from Fig.~\ref{ngprob}(a), the region of large $\mathcal{D^{\text{PS}}}$ corresponds to a low success probability.

For the Sym 1-PSTMSV state, the region of positive $\mathcal{D^{\text{PS}}}$ lies in a pocket of high transmissivity and low squeezing. As we subtract more photons, the size of the pocket increases. Again, the success probability for the corresponding positive $\mathcal{D^{\text{PS}}}$ region is low.

We can correlate these results with Figs.~\ref{phase_1d_l} and~\ref{phase_1d_t}. For instance, at $\tau=0.9$, Asym 1-PSTMSV yields a positive $\mathcal{D^{\text{PS}}}$ till $\lambda \approx 0.6$, which corroborates with Fig.~\ref{phase_1d_l}(a), where Asym 1-PSTMSV state crosses over the TMSV state at $\lambda \approx 0.6$.

\begin{figure}  
	\begin{center}
		\includegraphics[scale=1]{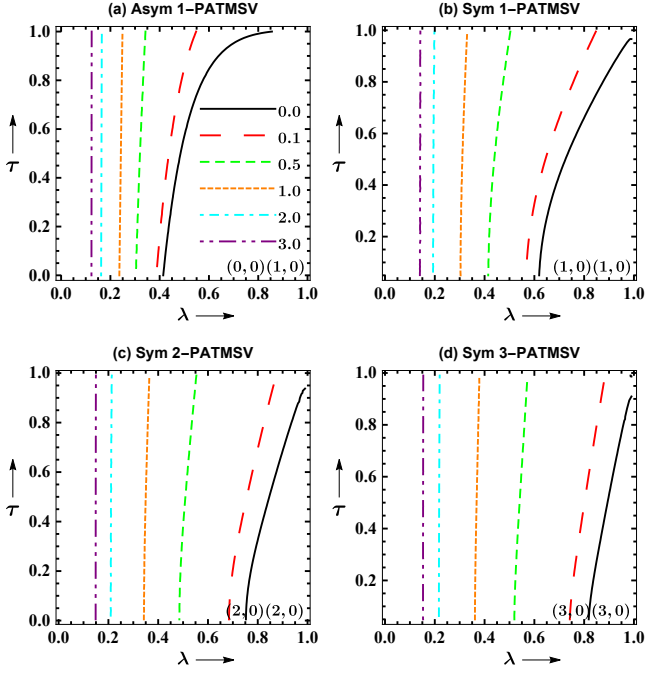}
		\caption{  Plots of   fixed $\mathcal{D^{\text{PA}}}$, difference of $\Delta \phi$ between TMSV and PATMSV states,  as a function of the transmissivity $\tau$ and
			squeezing parameter $\lambda$.   The value of $(m_1,n_1)(m_2,n_2)$ has been shown in the bottom right. We have set the phase $\phi=0.01$ for all the cases.
		}
		\label{phase_add_c}
	\end{center}
\end{figure}

We now plot various fixed values of $\mathcal{D^{\text{PA}}}$ as a function of the transmissivity $\tau$ and squeezing parameter $\lambda$ in Fig.~\ref{phase_add_c}. Since the expressions of $\Delta \phi$ for Asym 1-PSTMSV and Asym 1-PATMSV states are the same, the regions of positive $\mathcal{D^{\text{PS}}}$ and $\mathcal{D^{\text{PA}}}$ for these two states coincide.  For Sym 1-PATMSV state, we obtain positive $\mathcal{D^{\text{PA}}}$ region for even higher values of $\lambda$ as compared to Asym 1-PATMSV state. The region is further enlarged for a higher number of symmetric photon subtraction. Furthermore, we can see from Fig.~\ref{ngprob}(e)-(h), the positive $\mathcal{D^{\text{PA}}}$ region overlaps with a high success probability region.

\begin{figure} 
	\begin{center}
		\includegraphics[scale=1]{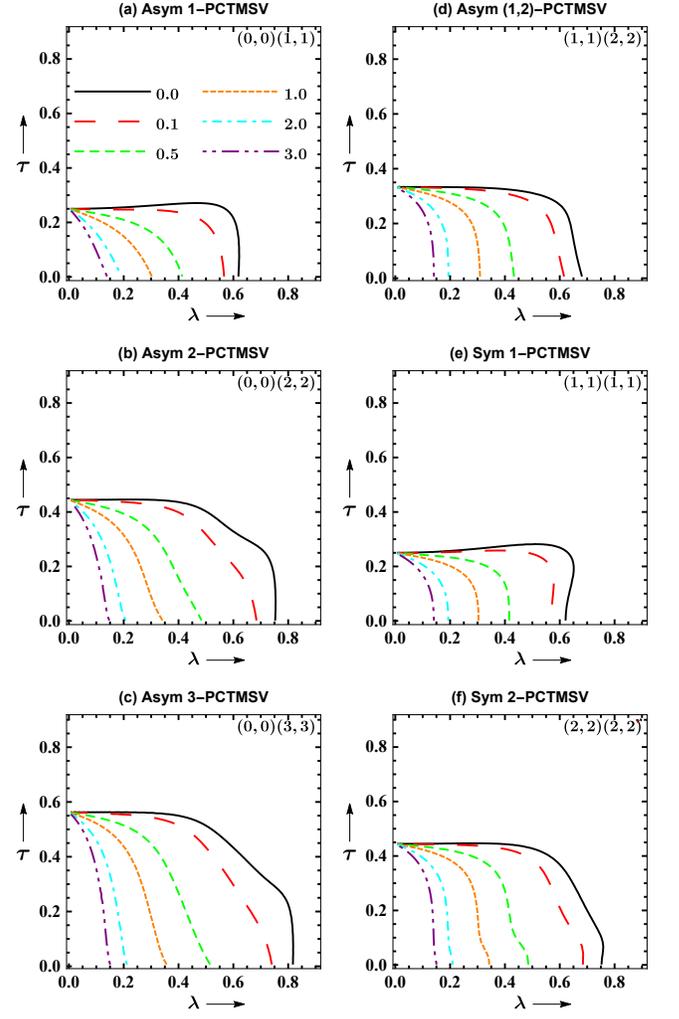}
		\caption{   Plots of    fixed $\mathcal{D^{\text{PC}}}$, difference of $\Delta \phi$ between TMSV and PCTMSV states,  as a function of the transmissivity $\tau$ and
			squeezing parameter $\lambda$. 
			The value of $(m_1,n_1)(m_2,n_2)$ has been shown in the top right. We have set the phase $\phi=0.01$ for all the cases.
		}
		\label{phase_cat_c}
	\end{center}
\end{figure}

Finally, we plot various fixed values of $\mathcal{D^{\text{PC}}}$ as a function of the transmissivity $\tau$ and squeezing parameter $\lambda$ in Fig.~\ref{phase_sub_c}. The results show that Asym $n$-PCTMSV and Sym $n$-PCTMSV states yield region with positive $\mathcal{D^{\text{PC}}}$ for all values of $n$. The region of positive $\mathcal{D^{\text{PC}}}$ lies in a pocket of low transmissivity and low squeezing for the Asym 1-PCTMSV and Sym 1-PCTMSV states. As we catalyze more photons asymmetrically and symmetrically, the size of the pocket increases. Here we have also considered the additional case of Asym (1,2)-PCTMSV state, where the catalysis of one and two photons is performed in modes $A_1$ and $A_2$ respectively, which yields positive result in low transmissivity and low squeezing regime.
The positive $\mathcal{D^{\text{PC}}}$  corresponds to a region of low success probability, as can be seen from Fig.~\ref{ngprob}(i)-(l).

 Next, we quantitatively take the success probability into account, where we aim to maximize the product 
$ P^{\text{NG}} \times \mathcal{D^{\text{NG}}}$.  More specifically, we intend to achieve an optimal trade-off
between   $P^{\text{NG}}$ and $\mathcal{D^{\text{NG}}}$  by
adjusting the transmissivity for a given squeezing.

\begin{figure}[H] 
	\begin{center}
		\includegraphics[scale=1]{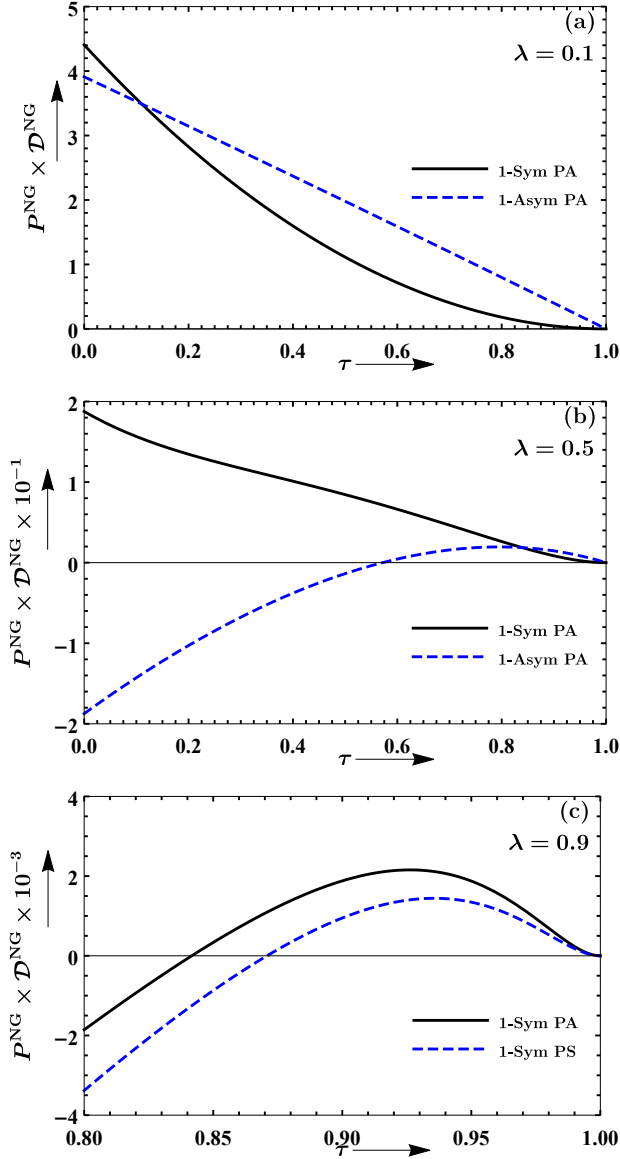}
		\caption{  Plot  of    $P^{\text{NG}} \times \mathcal{D^{\text{NG}}}  $  as a function of the transmissivity $\tau$ for different
			squeezing parameters. 
		  We have set the phase $\phi=0.01$ for all the cases.
		}
		\label{sensirate}
	\end{center}
\end{figure}

 We compare the value of the  product $ P^{\text{NG}} \times \mathcal{D^{\text{NG}}}$ for six different non-Gaussian operations namely 1-Asym PS, 1-Asym PA, 1-Asym PC, 1-Sym PS, 1-Sym PA, and 1-Sym PC. We plot $ P^{\text{NG}} \times \mathcal{D^{\text{NG}}}$ as a function of the transmissivity for different squeezing parameters in Fig.~\ref{sensirate}.
For panels (a) and (b), we have shown only those two curves, which provide maximum advantages for some values of transmissivity, whereas, for panel (c), the curves for two best performing non-Gaussian operations have been shown.
For small squeezing $\lambda=0.1$, 1-Sym PA operation outperforms all other operations in small transmissivity regions $0< \tau < 0.11$, whereas for all other values of transmissivity, 1-Asym operation PA maximizes the product. For intermediate squeezing $\lambda=0.5$, 1-Asym PA operation outperforms all other operations in high transmissivity regions $ 0.84 < \tau < 1 $, whereas for all other values of transmissivity, 1-Sym PA operation maximizes the product.
For high squeezing $\lambda=0.9$, 1-Asym PA operation outperforms all other operations in high transmissivity regions $ 0.84 < \tau < 1 $, whereas for all other values of transmissivity, $ P^{\text{NG}} \times \mathcal{D^{\text{NG}}}$ is negative for all non-Gaussian operation, and hence the TMSV state is superior to all other non-Gaussian states considered here.

To conclude this section, the phase sensitivity analysis, along with the success probability consideration, reveals that photon addition operation is the most advantageous among all the three non-Gaussian operations.  Specifically,  1-Sym PA operation in low transmissivity and squeezing regime provides maximum advantage.


\section{Conclusion}
\label{sec:conc}
In this paper, we derived the generalized Wigner function for non-Gaussian states, including PSTMSV, PATMSV, and PCTMSV states. The free parameters in the Wigner function include the squeezing parameter of the TMSV state and the transmissivity of the beam splitters used to implement the non-Gaussian operations. Further, one can choose the number of photons subtracted, added, or catalyzed on each of the modes of the TMSV state, as per requirement.
We then use this generalized Wigner function to calculate the lower bound on the phase sensitivity via QCRB and parity detection-based phase sensitivity in MZI.

We have considered a realistic photon subtraction and addition model, which yields the ideal case of photon subtraction and addition in the unit transmissivity limit. Therefore, the phase sensitivity results of ideal symmetric photon subtraction and addition~\cite{josab-2012,josab-2016}, as well as that of asymmetric catalysis~\cite{pra-catalysis-2021},  on TMSV state form a particular case of our work.

We also define a figure of merit as the difference between the phase sensitivity of the TMSV state and the NG-TMSV state, which enables us to identify the beneficial squeezing and transmissivity parameter range. Taking the success probability of non-Gaussian state generation into account, it turns out that the photon addition is the most useful operation   among all three non-Gaussian operations.

The current work clearly emphasizes the importance of the probabilistic nature of non-Gaussian state production in the phase sensitivity analysis.
We discuss several new avenues of future investigations briefly. As we have shown that multi-photon asymmetric subtraction and addition do not provide any advantage over TMSV state in phase estimation, it would be interesting to explore whether another measurement such as intensity difference in the two output modes of MZI improves the phase sensitivity for such states. The probabilistic nature of non-Gaussian operations should be considered while studying the effects of different measurements on phase sensitivity.
We have considered the implementation of photon addition using a beam splitter, which requires on-demand single-photon sources. Experimentally, photon addition is implemented using parametric down-conversion~\cite{Zavatta-2006, Zavatta-pra-2007}, and therefore, it is of immense importance to analyze the phase sensitivity in the parametric down-conversion-based photon addition model.

\section*{Acknowledgement}
	This paper is the first in a series of publications
written in the celebration of the completion of 15 years of IISER Mohali. We thank Profs. Arvind and Narayanasami Sathyamurthy for encouraging and enabling us to do independent research. C.K.  acknowledges the financial support from {\bf DST/ICPS/QuST/Theme-1/2019/General} Project number {\sf Q-68}.

\appendix
\begin{widetext}
	\section{Explicit form of the matrices in the Wigner function of the NG-TMSV state}\label{appwigner}
	Here we provide the explicit form of the matrices $M_1$, $M_2$, and $M_3$  which appear in the Wigner function of the NG-TMSV state~(\ref{eq4}). The matrix $M_1$ is given by
	\begin{equation}\label{mat1}
		M_1 = \frac{-1}{a_0}
		\left(
		\begin{array}{cccc}
			\alpha ^2 \left(t_1^2 t_2^2+1\right)+1 & 0 & -2 \alpha  \beta  t_1 t_2 & 0 \\
			0 & \alpha ^2 \left(t_1^2 t_2^2+1\right)+1 & 0 & 2 \alpha  \beta  t_1 t_2 \\
			-2 \alpha  \beta  t_1 t_2 & 0 & \alpha ^2 \left(t_1^2 t_2^2+1\right)+1 & 0 \\
			0 & 2 \alpha  \beta  t_1 t_2 & 0 & \alpha ^2 \left(t_1^2 t_2^2+1\right)+1 \\
		\end{array}
		\right),
	\end{equation}
	where $t_i=\sqrt{\tau_i}$ and $r_i=\sqrt{1-\tau_i}$ ($i=1,2$). Further, $\alpha=\sinh \, r$ and $\beta=\cosh \, r$. The matrix $M_2$ is given by
	\begin{equation}\label{mat2}
		M_2 = \frac{-1}{a_0}
		\left(
		\begin{array}{cccc}
			-\beta ^2 r_1 & -i \beta ^2 r_1 & \alpha  \beta  r_1 t_1 t_2 & -i \alpha  \beta  r_1 t_1 t_2 \\
			\beta ^2 r_1 & -i \beta ^2 r_1 & -\alpha  \beta  r_1 t_1 t_2 & -i \alpha  \beta  r_1 t_1 t_2 \\
			\alpha  \beta  r_2 t_1 t_2 & -i \alpha  \beta  r_2 t_1 t_2 & -\beta ^2 r_2 & -i \beta ^2 r_2 \\
			-\alpha  \beta  r_2 t_1 t_2 & -i \alpha  \beta  r_2 t_1 t_2 & \beta ^2 r_2 & -i \beta ^2 r_2 \\
			-\alpha ^2 r_1 t_1 t_2^2 & -i \alpha ^2 r_1 t_1 t_2^2 & \alpha  \beta  r_1 t_2 & -i \alpha  \beta  r_1 t_2 \\
			\alpha ^2 r_1 t_1 t_2^2 & -i \alpha ^2 r_1 t_1 t_2^2 & -\alpha  \beta  r_1 t_2 & -i \alpha  \beta  r_1 t_2 \\
			\alpha  \beta  r_2 t_1 & -i \alpha  \beta  r_2 t_1 & -\alpha ^2 r_2 t_1^2 t_2 & -i \alpha ^2 r_2 t_1^2 t_2 \\
			-\alpha  \beta  r_2 t_1 & -i \alpha  \beta  r_2 t_1 & \alpha ^2 r_2 t_1^2 t_2 & -i \alpha ^2 r_2 t_1^2 t_2 \\
		\end{array}
		\right),
	\end{equation}
	The matrix $M_3$ is given by
	{\small
		\begin{equation}\label{mat3}
			M_3 = \frac{-1}{4 a_0} 
			\left(
			\begin{array}{cccccccc}
				0 & -\beta ^2 r_1^2 & -\alpha  \beta  r_1 r_2 t_1 t_2 & 0 & 0 & \alpha ^2 r_2^2 t_1+t_1 & -\alpha  \beta  r_1 r_2 t_1 & 0 \\
				-\beta ^2 r_1^2 & 0 & 0 & -\alpha  \beta  r_1 r_2 t_1 t_2 & \alpha ^2 r_2^2 t_1+t_1 & 0 & 0 & -\alpha  \beta  r_1 r_2 t_1 \\
				-\alpha  \beta  r_1 r_2 t_1 t_2 & 0 & 0 & -\beta ^2 r_2^2 & -\alpha  \beta  r_1 r_2 t_2 & 0 & 0 & \alpha ^2 r_1^2 t_2+t_2 \\
				0 & -\alpha  \beta  r_1 r_2 t_1 t_2 & -\beta ^2 r_2^2 & 0 & 0 & -\alpha  \beta  r_1 r_2 t_2 & \alpha ^2 r_1^2 t_2+t_2 & 0 \\
				0 & \alpha ^2 r_2^2 t_1+t_1 & -\alpha  \beta  r_1 r_2 t_2 & 0 & 0 & -\alpha ^2 r_1^2 t_2^2 & -\alpha  \beta  r_1 r_2 & 0 \\
				\alpha ^2 r_2^2 t_1+t_1 & 0 & 0 & -\alpha  \beta  r_1 r_2 t_2 & -\alpha ^2 r_1^2 t_2^2 & 0 & 0 & -\alpha  \beta  r_1 r_2 \\
				-\alpha  \beta  r_1 r_2 t_1 & 0 & 0 & \alpha ^2 r_1^2 t_2+t_2 & -\alpha  \beta  r_1 r_2 & 0 & 0 & -\alpha ^2 r_2^2 t_1^2 \\
				0 & -\alpha  \beta  r_1 r_2 t_1 & \alpha ^2 r_1^2 t_2+t_2 & 0 & 0 & -\alpha  \beta  r_1 r_2 & -\alpha ^2 r_2^2 t_1^2 & 0 \\
			\end{array}
			\right).
		\end{equation}
	}
	\section{Explicit form of the matrix in the probability expression}\label{appprob}
	The matrix $M_4$ appearing in the success probability expression~(\ref{successp}) is given by
	{\small
		\begin{equation}\label{mat4}
			M_4 = \frac{-1}{4 a_0}  \left(
			\begin{array}{cccccccc}
				0 & \beta ^2 r_1^2 & -\alpha  \beta  r_1 r_2 t_1 t_2 & 0 & 0 & \alpha ^2 r_2^2 t_1+t_1 & \alpha  \beta  r_1 r_2 t_1 & 0 \\
				\beta ^2 r_1^2 & 0 & 0 & -\alpha  \beta  r_1 r_2 t_1 t_2 & \alpha ^2 r_2^2 t_1+t_1 & 0 & 0 & \alpha  \beta  r_1 r_2 t_1 \\
				-\alpha  \beta  r_1 r_2 t_1 t_2 & 0 & 0 & \beta ^2 r_2^2 & \alpha  \beta  r_1 r_2 t_2 & 0 & 0 & \alpha ^2 r_1^2 t_2+t_2 \\
				0 & -\alpha  \beta  r_1 r_2 t_1 t_2 & \beta ^2 r_2^2 & 0 & 0 & \alpha  \beta  r_1 r_2 t_2 & \alpha ^2 r_1^2 t_2+t_2 & 0 \\
				0 & \alpha ^2 r_2^2 t_1+t_1 & \alpha  \beta  r_1 r_2 t_2 & 0 & 0 & \alpha ^2 r_1^2 t_2^2 & -\alpha  \beta  r_1 r_2 & 0 \\
				\alpha ^2 r_2^2 t_1+t_1 & 0 & 0 & \alpha  \beta  r_1 r_2 t_2 & \alpha ^2 r_1^2 t_2^2 & 0 & 0 & -\alpha  \beta  r_1 r_2 \\
				\alpha  \beta  r_1 r_2 t_1 & 0 & 0 & \alpha ^2 r_1^2 t_2+t_2 & -\alpha  \beta  r_1 r_2 & 0 & 0 & \alpha ^2 r_2^2 t_1^2 \\
				0 & \alpha  \beta  r_1 r_2 t_1 & \alpha ^2 r_1^2 t_2+t_2 & 0 & 0 & -\alpha  \beta  r_1 r_2 & \alpha ^2 r_2^2 t_1^2 & 0 \\
			\end{array}
			\right).
		\end{equation}
	}
	\section{Explicit form of the matrices in the moment generating function}\label{appmomgen}
	The matrix $M_4$ arising in the expression of the moment generating function~(\ref{momgen}) has already been given  in Eq.~(\ref{mat4}), we now provide the matrices $M_5$ and $M_6$:
	\begin{equation}\label{mat5}
		M_5=\frac{-1}{2 a_0} \left(
		\begin{array}{cccc}
			-\beta ^2 r_1 & -i \beta ^2 r_1 & -\alpha  \beta  r_1 t_1 t_2 & i \alpha  \beta  r_1 t_1 t_2 \\
			\beta ^2 r_1 & -i \beta ^2 r_1 & \alpha  \beta  r_1 t_1 t_2 & i \alpha  \beta  r_1 t_1 t_2 \\
			-\alpha  \beta  r_2 t_1 t_2 & i \alpha  \beta  r_2 t_1 t_2 & -\beta ^2 r_2 & -i \beta ^2 r_2 \\
			\alpha  \beta  r_2 t_1 t_2 & i \alpha  \beta  r_2 t_1 t_2 & \beta ^2 r_2 & -i \beta ^2 r_2 \\
			\alpha ^2 r_1 t_1 t_2^2 & i \alpha ^2 r_1 t_1 t_2^2 & \alpha  \beta  r_1 t_2 & -i \alpha  \beta  r_1 t_2 \\
			-\alpha ^2 r_1 t_1 t_2^2 & i \alpha ^2 r_1 t_1 t_2^2 & -\alpha  \beta  r_1 t_2 & -i \alpha  \beta  r_1 t_2 \\
			\alpha  \beta  r_2 t_1 & -i \alpha  \beta  r_2 t_1 & \alpha ^2 r_2 t_1^2 t_2 & i \alpha ^2 r_2 t_1^2 t_2 \\
			-\alpha  \beta  r_2 t_1 & -i \alpha  \beta  r_2 t_1 & -\alpha ^2 r_2 t_1^2 t_2 & i \alpha ^2 r_2 t_1^2 t_2 \\
		\end{array}
		\right),
	\end{equation}
	and
	\begin{equation}\label{mat6}
		M_6=\frac{1}{4 a_0} \left(
		\begin{array}{cccc}
			\alpha ^2 \left(t_1^2 t_2^2+1\right)+1 & 0 & 2 \alpha  \beta  t_1 t_2 & 0 \\
			0 & \alpha ^2 \left(t_1^2 t_2^2+1\right)+1 & 0 & -2 \alpha  \beta  t_1 t_2 \\
			2 \alpha  \beta  t_1 t_2 & 0 & \alpha ^2 \left(t_1^2 t_2^2+1\right)+1 & 0 \\
			0 & -2 \alpha  \beta  t_1 t_2 & 0 & \alpha ^2 \left(t_1^2 t_2^2+1\right)+1 \\
		\end{array}
		\right).
	\end{equation}
	\section{Matrix in the expectation of the parity operator}\label{apppari}
	The matrix $M_7$ appearing in the average of the parity operator~(\ref{apari}) is given by
	\begin{equation}\label{mat7}
		M_7=\frac{-1}{4 w_0} \left(
		\begin{array}{cccccccc}
			w_1 & w_2 & w_3 & w_4 & w_5 & w_6 & w_7 & w_8 \\
			w_2 & w_1 & w_4 & w_3 & w_6 & w_5 & w_8 & w_7 \\
			w_3 & w_4 & w_9 & w_{10} & w_{11} & w_{12} & w_{13} & w_{14} \\
			w_4 & w_3 & w_{10} & w_9 & w_{12} & w_{11} & w_{14} & w_{13} \\
			w_5 & w_6 & w_{11} & w_{12} & w_{15} & w_{16} & w_{17} & w_{18} \\
			w_6 & w_5 & w_{12} & w_{11} & w_{16} & w_{15} & w_{18} & w_{17} \\
			w_7 & w_8 & w_{13} & w_{14} & w_{17} & w_{18} & w_{19} & w_{20} \\
			w_8 & w_7 & w_{14} & w_{13} & w_{18} & w_{17} & w_{20} & w_{19} \\
		\end{array}
		\right),
	\end{equation}
	where
	\begin{equation}
			\begin{array}{ccccccc}
				w_0&=& 2 c_2 \lambda ^2 t_1^2 t_2^2+\lambda ^4 t_1^4 t_2^4+1, & \qquad & w_{11} & = & -c_1 \lambda r_1 r_2 t_2 \left(\lambda ^2 t_1^2 t_2^2+1\right), \\
				w_1 & = & \lambda  r_1^2 s_2 t_1 t_2, & \qquad & w_{12} & = & -2 c_1 \lambda ^2 r_1 r_2 s_1 t_1 t_2^2, \\
				w_2 & = & c_1 r_1^2 \left(\lambda ^2 t_1^2 t_2^2+1\right), & \qquad & w_{13} & = & \lambda  r_2^2 s_1 t_1 \left(\lambda ^2 t_1^2 t_2^2-1\right), \\
				w_3 & = & \lambda  r_1 r_2 t_1 t_2 \left(c_2+\lambda ^2 t_1^2 t_2^2\right), & \qquad & w_{14} & = & \lambda ^2 t_1^2 t_2^3 \left(c_2+\lambda ^2 t_1^2\right)+c_2 \lambda ^2 t_1^2 t_2+t_2, \\
				w_4 & = & r_1 r_2 s_1 \left(\lambda ^2 t_1^2 t_2^2-1\right), & \qquad & w_{15} & = & -\lambda ^3 r_1^2 s_2 t_1 t_2^3, \\
				w_5 & = & \lambda  r_1^2 s_1 t_2 \left(\lambda ^2 t_1^2 t_2^2-1\right), & \qquad & w_{16} & = & -c_1 \lambda^2 r_1^2 t_2^2 \left(\lambda ^2 t_1^2 t_2^2+1\right), \\
				w_6 & = & \lambda ^2 t_2^2 t_1^3 \left(c_2+\lambda ^2 t_2^2\right)+c_2 \lambda ^2 t_2^2 t_1+t_1, & \qquad & w_{17} & = & -\lambda  r_1 r_2 \left(c_2 \lambda ^2 t_1^2 t_2^2+1\right), \\
				w_7 & = & c_1 \lambda  r_1 r_2 t_1 \left(\lambda ^2 t_1^2 t_2^2+1\right), & \qquad & w_{18} & = & \lambda ^2 r_1 r_2 s_1 t_1 t_2 \left(\lambda ^2 t_1^2 t_2^2-1\right), \\
				w_8 & = & 2 c_1 \lambda ^2 r_1 r_2 s_1 t_1^2 t_2, & \qquad & w_{19} & = & \lambda ^3 r_2^2 s_2 t_1^3 t_2, \\
				w_9 & = & -2 c_1 \lambda  r_2^2 s_1 t_1 t_2, & \qquad & w_{20} & = & c_1 \lambda ^2 r_2^2 t_1^2 \left(\lambda ^2 t_1^2 t_2^2+1\right), \\
				w_{10} & = & -c_1 r_2^2 \left(\lambda ^2 t_1^2 t_2^2+1\right), &&&&     \\
			\end{array}
		\end{equation}	
		with $c_1=\cos  \, \phi $, $s_1=\sin \, \phi $, $c_2=\cos (2\phi)$, and $s_2=\sin (2\phi)$.  
\end{widetext}

\begin{thebibliography}{53}%
	\makeatletter
	\providecommand \@ifxundefined [1]{%
		\@ifx{#1\undefined}
	}%
	\providecommand \@ifnum [1]{%
		\ifnum #1\expandafter \@firstoftwo
		\else \expandafter \@secondoftwo
		\fi
	}%
	\providecommand \@ifx [1]{%
		\ifx #1\expandafter \@firstoftwo
		\else \expandafter \@secondoftwo
		\fi
	}%
	\providecommand \natexlab [1]{#1}%
	\providecommand \enquote  [1]{``#1''}%
	\providecommand \bibnamefont  [1]{#1}%
	\providecommand \bibfnamefont [1]{#1}%
	\providecommand \citenamefont [1]{#1}%
	\providecommand \href@noop [0]{\@secondoftwo}%
	\providecommand \href [0]{\begingroup \@sanitize@url \@href}%
	\providecommand \@href[1]{\@@startlink{#1}\@@href}%
	\providecommand \@@href[1]{\endgroup#1\@@endlink}%
	\providecommand \@sanitize@url [0]{\catcode `\\12\catcode `\$12\catcode
		`\&12\catcode `\#12\catcode `\^12\catcode `\_12\catcode `\%12\relax}%
	\providecommand \@@startlink[1]{}%
	\providecommand \@@endlink[0]{}%
	\providecommand \url  [0]{\begingroup\@sanitize@url \@url }%
	\providecommand \@url [1]{\endgroup\@href {#1}{\urlprefix }}%
	\providecommand \urlprefix  [0]{URL }%
	\providecommand \Eprint [0]{\href }%
	\providecommand \doibase [0]{http://dx.doi.org/}%
	\providecommand \selectlanguage [0]{\@gobble}%
	\providecommand \bibinfo  [0]{\@secondoftwo}%
	\providecommand \bibfield  [0]{\@secondoftwo}%
	\providecommand \translation [1]{[#1]}%
	\providecommand \BibitemOpen [0]{}%
	\providecommand \bibitemStop [0]{}%
	\providecommand \bibitemNoStop [0]{.\EOS\space}%
	\providecommand \EOS [0]{\spacefactor3000\relax}%
	\providecommand \BibitemShut  [1]{\csname bibitem#1\endcsname}%
	\let\auto@bib@innerbib\@empty
	\bibitem [{\citenamefont {Dowling}(2008)}]{Dowling-cp-2008}%
	\BibitemOpen
	\bibfield  {author} {\bibinfo {author} {\bibfnamefont {J.~P.}\ \bibnamefont
			{Dowling}},\ }\href {\doibase 10.1080/00107510802091298} {\bibfield
		{journal} {\bibinfo  {journal} {Contemporary Physics}\ }\textbf {\bibinfo
			{volume} {49}},\ \bibinfo {pages} {125} (\bibinfo {year} {2008})}\BibitemShut
	{NoStop}%
	\bibitem [{\citenamefont {Giovannetti}\ \emph {et~al.}(2011)\citenamefont
		{Giovannetti}, \citenamefont {Lloyd},\ and\ \citenamefont
		{Maccone}}]{Giovannetti2011}%
	\BibitemOpen
	\bibfield  {author} {\bibinfo {author} {\bibfnamefont {V.}~\bibnamefont
			{Giovannetti}}, \bibinfo {author} {\bibfnamefont {S.}~\bibnamefont {Lloyd}},
		\ and\ \bibinfo {author} {\bibfnamefont {L.}~\bibnamefont {Maccone}},\ }\href
	{\doibase 10.1038/nphoton.2011.35} {\bibfield  {journal} {\bibinfo  {journal}
			{Nature Photonics}\ }\textbf {\bibinfo {volume} {5}},\ \bibinfo {pages} {222}
		(\bibinfo {year} {2011})}\BibitemShut {NoStop}%
	\bibitem [{\citenamefont {Caves}(1981)}]{caves-prd-1981}%
	\BibitemOpen
	\bibfield  {author} {\bibinfo {author} {\bibfnamefont {C.~M.}\ \bibnamefont
			{Caves}},\ }\href {\doibase 10.1103/PhysRevD.23.1693} {\bibfield  {journal}
		{\bibinfo  {journal} {Phys. Rev. D}\ }\textbf {\bibinfo {volume} {23}},\
		\bibinfo {pages} {1693} (\bibinfo {year} {1981})}\BibitemShut {NoStop}%
	\bibitem [{\citenamefont {Kwon}\ \emph {et~al.}(2019)\citenamefont {Kwon},
		\citenamefont {Tan}, \citenamefont {Volkoff},\ and\ \citenamefont
		{Jeong}}]{Jeong-prl-2019}%
	\BibitemOpen
	\bibfield  {author} {\bibinfo {author} {\bibfnamefont {H.}~\bibnamefont
			{Kwon}}, \bibinfo {author} {\bibfnamefont {K.~C.}\ \bibnamefont {Tan}},
		\bibinfo {author} {\bibfnamefont {T.}~\bibnamefont {Volkoff}}, \ and\
		\bibinfo {author} {\bibfnamefont {H.}~\bibnamefont {Jeong}},\ }\href
	{\doibase 10.1103/PhysRevLett.122.040503} {\bibfield  {journal} {\bibinfo
			{journal} {Phys. Rev. Lett.}\ }\textbf {\bibinfo {volume} {122}},\ \bibinfo
		{pages} {040503} (\bibinfo {year} {2019})}\BibitemShut {NoStop}%
	\bibitem [{\citenamefont {Hofmann}\ and\ \citenamefont
		{Ono}(2007)}]{Hofmann-pra-2007}%
	\BibitemOpen
	\bibfield  {author} {\bibinfo {author} {\bibfnamefont {H.~F.}\ \bibnamefont
			{Hofmann}}\ and\ \bibinfo {author} {\bibfnamefont {T.}~\bibnamefont {Ono}},\
	}\href {\doibase 10.1103/PhysRevA.76.031806} {\bibfield  {journal} {\bibinfo
			{journal} {Phys. Rev. A}\ }\textbf {\bibinfo {volume} {76}},\ \bibinfo
		{pages} {031806} (\bibinfo {year} {2007})}\BibitemShut {NoStop}%
	\bibitem [{\citenamefont {Anisimov}\ \emph {et~al.}(2010)\citenamefont
		{Anisimov}, \citenamefont {Raterman}, \citenamefont {Chiruvelli},
		\citenamefont {Plick}, \citenamefont {Huver}, \citenamefont {Lee},\ and\
		\citenamefont {Dowling}}]{Anisimov-prl-2010}%
	\BibitemOpen
	\bibfield  {author} {\bibinfo {author} {\bibfnamefont {P.~M.}\ \bibnamefont
			{Anisimov}}, \bibinfo {author} {\bibfnamefont {G.~M.}\ \bibnamefont
			{Raterman}}, \bibinfo {author} {\bibfnamefont {A.}~\bibnamefont
			{Chiruvelli}}, \bibinfo {author} {\bibfnamefont {W.~N.}\ \bibnamefont
			{Plick}}, \bibinfo {author} {\bibfnamefont {S.~D.}\ \bibnamefont {Huver}},
		\bibinfo {author} {\bibfnamefont {H.}~\bibnamefont {Lee}}, \ and\ \bibinfo
		{author} {\bibfnamefont {J.~P.}\ \bibnamefont {Dowling}},\ }\href {\doibase
		10.1103/PhysRevLett.104.103602} {\bibfield  {journal} {\bibinfo  {journal}
			{Phys. Rev. Lett.}\ }\textbf {\bibinfo {volume} {104}},\ \bibinfo {pages}
		{103602} (\bibinfo {year} {2010})}\BibitemShut {NoStop}%
	\bibitem [{\citenamefont {Giovannetti}\ \emph {et~al.}(2004)\citenamefont
		{Giovannetti}, \citenamefont {Lloyd},\ and\ \citenamefont
		{Maccone}}]{Giovannetti-science-2004}%
	\BibitemOpen
	\bibfield  {author} {\bibinfo {author} {\bibfnamefont {V.}~\bibnamefont
			{Giovannetti}}, \bibinfo {author} {\bibfnamefont {S.}~\bibnamefont {Lloyd}},
		\ and\ \bibinfo {author} {\bibfnamefont {L.}~\bibnamefont {Maccone}},\ }\href
	{\doibase 10.1126/science.1104149} {\bibfield  {journal} {\bibinfo  {journal}
			{Science}\ }\textbf {\bibinfo {volume} {306}},\ \bibinfo {pages} {1330}
		(\bibinfo {year} {2004})}\BibitemShut {NoStop}%
	\bibitem [{\citenamefont {Gerry}(2000)}]{gerry-pra-2000}%
	\BibitemOpen
	\bibfield  {author} {\bibinfo {author} {\bibfnamefont {C.~C.}\ \bibnamefont
			{Gerry}},\ }\href {\doibase 10.1103/PhysRevA.61.043811} {\bibfield  {journal}
		{\bibinfo  {journal} {Phys. Rev. A}\ }\textbf {\bibinfo {volume} {61}},\
		\bibinfo {pages} {043811} (\bibinfo {year} {2000})}\BibitemShut {NoStop}%
	\bibitem [{\citenamefont {Gerry}\ and\ \citenamefont
		{Campos}(2001)}]{gerry-pra-2001}%
	\BibitemOpen
	\bibfield  {author} {\bibinfo {author} {\bibfnamefont {C.~C.}\ \bibnamefont
			{Gerry}}\ and\ \bibinfo {author} {\bibfnamefont {R.~A.}\ \bibnamefont
			{Campos}},\ }\href {\doibase 10.1103/PhysRevA.64.063814} {\bibfield
		{journal} {\bibinfo  {journal} {Phys. Rev. A}\ }\textbf {\bibinfo {volume}
			{64}},\ \bibinfo {pages} {063814} (\bibinfo {year} {2001})}\BibitemShut
	{NoStop}%
	\bibitem [{\citenamefont {Gerry}\ and\ \citenamefont
		{Benmoussa}(2002)}]{gerry-pra-2002}%
	\BibitemOpen
	\bibfield  {author} {\bibinfo {author} {\bibfnamefont {C.~C.}\ \bibnamefont
			{Gerry}}\ and\ \bibinfo {author} {\bibfnamefont {A.}~\bibnamefont
			{Benmoussa}},\ }\href {\doibase 10.1103/PhysRevA.65.033822} {\bibfield
		{journal} {\bibinfo  {journal} {Phys. Rev. A}\ }\textbf {\bibinfo {volume}
			{65}},\ \bibinfo {pages} {033822} (\bibinfo {year} {2002})}\BibitemShut
	{NoStop}%
	\bibitem [{\citenamefont {Campos}\ \emph {et~al.}(2003)\citenamefont {Campos},
		\citenamefont {Gerry},\ and\ \citenamefont {Benmoussa}}]{gerry-pra-2003}%
	\BibitemOpen
	\bibfield  {author} {\bibinfo {author} {\bibfnamefont {R.~A.}\ \bibnamefont
			{Campos}}, \bibinfo {author} {\bibfnamefont {C.~C.}\ \bibnamefont {Gerry}}, \
		and\ \bibinfo {author} {\bibfnamefont {A.}~\bibnamefont {Benmoussa}},\ }\href
	{\doibase 10.1103/PhysRevA.68.023810} {\bibfield  {journal} {\bibinfo
			{journal} {Phys. Rev. A}\ }\textbf {\bibinfo {volume} {68}},\ \bibinfo
		{pages} {023810} (\bibinfo {year} {2003})}\BibitemShut {NoStop}%
	\bibitem [{\citenamefont {Gerry}\ and\ \citenamefont
		{Mimih}(2010{\natexlab{a}})}]{gerry-pra-2010}%
	\BibitemOpen
	\bibfield  {author} {\bibinfo {author} {\bibfnamefont {C.~C.}\ \bibnamefont
			{Gerry}}\ and\ \bibinfo {author} {\bibfnamefont {J.}~\bibnamefont {Mimih}},\
	}\href {\doibase 10.1103/PhysRevA.82.013831} {\bibfield  {journal} {\bibinfo
			{journal} {Phys. Rev. A}\ }\textbf {\bibinfo {volume} {82}},\ \bibinfo
		{pages} {013831} (\bibinfo {year} {2010}{\natexlab{a}})}\BibitemShut
	{NoStop}%
	\bibitem [{\citenamefont {Gerry}\ and\ \citenamefont
		{Mimih}(2010{\natexlab{b}})}]{gerry-cp-2010}%
	\BibitemOpen
	\bibfield  {author} {\bibinfo {author} {\bibfnamefont {C.~C.}\ \bibnamefont
			{Gerry}}\ and\ \bibinfo {author} {\bibfnamefont {J.}~\bibnamefont {Mimih}},\
	}\href {\doibase 10.1080/00107514.2010.509995} {\bibfield  {journal}
		{\bibinfo  {journal} {Contemporary Physics}\ }\textbf {\bibinfo {volume}
			{51}},\ \bibinfo {pages} {497} (\bibinfo {year}
		{2010}{\natexlab{b}})}\BibitemShut {NoStop}%
	\bibitem [{\citenamefont {Joo}\ \emph {et~al.}(2011)\citenamefont {Joo},
		\citenamefont {Munro},\ and\ \citenamefont {Spiller}}]{ecs-prl-2011}%
	\BibitemOpen
	\bibfield  {author} {\bibinfo {author} {\bibfnamefont {J.}~\bibnamefont
			{Joo}}, \bibinfo {author} {\bibfnamefont {W.~J.}\ \bibnamefont {Munro}}, \
		and\ \bibinfo {author} {\bibfnamefont {T.~P.}\ \bibnamefont {Spiller}},\
	}\href {\doibase 10.1103/PhysRevLett.107.083601} {\bibfield  {journal}
		{\bibinfo  {journal} {Phys. Rev. Lett.}\ }\textbf {\bibinfo {volume} {107}},\
		\bibinfo {pages} {083601} (\bibinfo {year} {2011})}\BibitemShut {NoStop}%
	\bibitem [{\citenamefont {Seshadreesan}\ \emph {et~al.}(2011)\citenamefont
		{Seshadreesan}, \citenamefont {Anisimov}, \citenamefont {Lee},\ and\
		\citenamefont {Dowling}}]{Seshadreesan_2011}%
	\BibitemOpen
	\bibfield  {author} {\bibinfo {author} {\bibfnamefont {K.~P.}\ \bibnamefont
			{Seshadreesan}}, \bibinfo {author} {\bibfnamefont {P.~M.}\ \bibnamefont
			{Anisimov}}, \bibinfo {author} {\bibfnamefont {H.}~\bibnamefont {Lee}}, \
		and\ \bibinfo {author} {\bibfnamefont {J.~P.}\ \bibnamefont {Dowling}},\
	}\href {\doibase 10.1088/1367-2630/13/8/083026} {\bibfield  {journal}
		{\bibinfo  {journal} {New Journal of Physics}\ }\textbf {\bibinfo {volume}
			{13}},\ \bibinfo {pages} {083026} (\bibinfo {year} {2011})}\BibitemShut
	{NoStop}%
	\bibitem [{\citenamefont {Plick}\ \emph {et~al.}(2010)\citenamefont {Plick},
		\citenamefont {Anisimov}, \citenamefont {Dowling}, \citenamefont {Lee},\ and\
		\citenamefont {Agarwal}}]{Plick_2010}%
	\BibitemOpen
	\bibfield  {author} {\bibinfo {author} {\bibfnamefont {W.~N.}\ \bibnamefont
			{Plick}}, \bibinfo {author} {\bibfnamefont {P.~M.}\ \bibnamefont {Anisimov}},
		\bibinfo {author} {\bibfnamefont {J.~P.}\ \bibnamefont {Dowling}}, \bibinfo
		{author} {\bibfnamefont {H.}~\bibnamefont {Lee}}, \ and\ \bibinfo {author}
		{\bibfnamefont {G.~S.}\ \bibnamefont {Agarwal}},\ }\href {\doibase
		10.1088/1367-2630/12/11/113025} {\bibfield  {journal} {\bibinfo  {journal}
			{New Journal of Physics}\ }\textbf {\bibinfo {volume} {12}},\ \bibinfo
		{pages} {113025} (\bibinfo {year} {2010})}\BibitemShut {NoStop}%
	\bibitem [{\citenamefont {Chiruvelli}\ and\ \citenamefont
		{Lee}(2011)}]{aravind-2011}%
	\BibitemOpen
	\bibfield  {author} {\bibinfo {author} {\bibfnamefont {A.}~\bibnamefont
			{Chiruvelli}}\ and\ \bibinfo {author} {\bibfnamefont {H.}~\bibnamefont
			{Lee}},\ }\href {\doibase 10.1080/09500340.2011.585251} {\bibfield  {journal}
		{\bibinfo  {journal} {Journal of Modern Optics}\ }\textbf {\bibinfo {volume}
			{58}},\ \bibinfo {pages} {945} (\bibinfo {year} {2011})}\BibitemShut
	{NoStop}%
	\bibitem [{\citenamefont {Seshadreesan}\ \emph {et~al.}(2013)\citenamefont
		{Seshadreesan}, \citenamefont {Kim}, \citenamefont {Dowling},\ and\
		\citenamefont {Lee}}]{sesha-pra-2013}%
	\BibitemOpen
	\bibfield  {author} {\bibinfo {author} {\bibfnamefont {K.~P.}\ \bibnamefont
			{Seshadreesan}}, \bibinfo {author} {\bibfnamefont {S.}~\bibnamefont {Kim}},
		\bibinfo {author} {\bibfnamefont {J.~P.}\ \bibnamefont {Dowling}}, \ and\
		\bibinfo {author} {\bibfnamefont {H.}~\bibnamefont {Lee}},\ }\href {\doibase
		10.1103/PhysRevA.87.043833} {\bibfield  {journal} {\bibinfo  {journal} {Phys.
				Rev. A}\ }\textbf {\bibinfo {volume} {87}},\ \bibinfo {pages} {043833}
		(\bibinfo {year} {2013})}\BibitemShut {NoStop}%
	\bibitem [{\citenamefont {Sahota}\ and\ \citenamefont
		{James}(2013)}]{sahota-pra-2013}%
	\BibitemOpen
	\bibfield  {author} {\bibinfo {author} {\bibfnamefont {J.}~\bibnamefont
			{Sahota}}\ and\ \bibinfo {author} {\bibfnamefont {D.~F.~V.}\ \bibnamefont
			{James}},\ }\href {\doibase 10.1103/PhysRevA.88.063820} {\bibfield  {journal}
		{\bibinfo  {journal} {Phys. Rev. A}\ }\textbf {\bibinfo {volume} {88}},\
		\bibinfo {pages} {063820} (\bibinfo {year} {2013})}\BibitemShut {NoStop}%
	\bibitem [{\citenamefont {Zhang}\ \emph {et~al.}(2013)\citenamefont {Zhang},
		\citenamefont {Yang},\ and\ \citenamefont {Wang}}]{zhang-pra-2013}%
	\BibitemOpen
	\bibfield  {author} {\bibinfo {author} {\bibfnamefont {X.-X.}\ \bibnamefont
			{Zhang}}, \bibinfo {author} {\bibfnamefont {Y.-X.}\ \bibnamefont {Yang}}, \
		and\ \bibinfo {author} {\bibfnamefont {X.-B.}\ \bibnamefont {Wang}},\ }\href
	{\doibase 10.1103/PhysRevA.88.013838} {\bibfield  {journal} {\bibinfo
			{journal} {Phys. Rev. A}\ }\textbf {\bibinfo {volume} {88}},\ \bibinfo
		{pages} {013838} (\bibinfo {year} {2013})}\BibitemShut {NoStop}%
	\bibitem [{\citenamefont {Huver}\ \emph {et~al.}(2008)\citenamefont {Huver},
		\citenamefont {Wildfeuer},\ and\ \citenamefont {Dowling}}]{Dowling-pra-2008}%
	\BibitemOpen
	\bibfield  {author} {\bibinfo {author} {\bibfnamefont {S.~D.}\ \bibnamefont
			{Huver}}, \bibinfo {author} {\bibfnamefont {C.~F.}\ \bibnamefont
			{Wildfeuer}}, \ and\ \bibinfo {author} {\bibfnamefont {J.~P.}\ \bibnamefont
			{Dowling}},\ }\href {\doibase 10.1103/PhysRevA.78.063828} {\bibfield
		{journal} {\bibinfo  {journal} {Phys. Rev. A}\ }\textbf {\bibinfo {volume}
			{78}},\ \bibinfo {pages} {063828} (\bibinfo {year} {2008})}\BibitemShut
	{NoStop}%
	\bibitem [{\citenamefont {Vahlbruch}\ \emph {et~al.}(2016)\citenamefont
		{Vahlbruch}, \citenamefont {Mehmet}, \citenamefont {Danzmann},\ and\
		\citenamefont {Schnabel}}]{15dB}%
	\BibitemOpen
	\bibfield  {author} {\bibinfo {author} {\bibfnamefont {H.}~\bibnamefont
			{Vahlbruch}}, \bibinfo {author} {\bibfnamefont {M.}~\bibnamefont {Mehmet}},
		\bibinfo {author} {\bibfnamefont {K.}~\bibnamefont {Danzmann}}, \ and\
		\bibinfo {author} {\bibfnamefont {R.}~\bibnamefont {Schnabel}},\ }\href
	{\doibase 10.1103/PhysRevLett.117.110801} {\bibfield  {journal} {\bibinfo
			{journal} {Phys. Rev. Lett.}\ }\textbf {\bibinfo {volume} {117}},\ \bibinfo
		{pages} {110801} (\bibinfo {year} {2016})}\BibitemShut {NoStop}%
	\bibitem [{\citenamefont {Opatrn\'y}\ \emph {et~al.}(2000)\citenamefont
		{Opatrn\'y}, \citenamefont {Kurizki},\ and\ \citenamefont
		{Welsch}}]{tel2000}%
	\BibitemOpen
	\bibfield  {author} {\bibinfo {author} {\bibfnamefont {T.}~\bibnamefont
			{Opatrn\'y}}, \bibinfo {author} {\bibfnamefont {G.}~\bibnamefont {Kurizki}},
		\ and\ \bibinfo {author} {\bibfnamefont {D.-G.}\ \bibnamefont {Welsch}},\
	}\href {\doibase 10.1103/PhysRevA.61.032302} {\bibfield  {journal} {\bibinfo
			{journal} {Phys. Rev. A}\ }\textbf {\bibinfo {volume} {61}},\ \bibinfo
		{pages} {032302} (\bibinfo {year} {2000})}\BibitemShut {NoStop}%
	\bibitem [{\citenamefont {Yang}\ and\ \citenamefont {Li}(2009)}]{tel2009}%
	\BibitemOpen
	\bibfield  {author} {\bibinfo {author} {\bibfnamefont {Y.}~\bibnamefont
			{Yang}}\ and\ \bibinfo {author} {\bibfnamefont {F.-L.}\ \bibnamefont {Li}},\
	}\href {\doibase 10.1103/PhysRevA.80.022315} {\bibfield  {journal} {\bibinfo
			{journal} {Phys. Rev. A}\ }\textbf {\bibinfo {volume} {80}},\ \bibinfo
		{pages} {022315} (\bibinfo {year} {2009})}\BibitemShut {NoStop}%
	\bibitem [{\citenamefont {Xu}(2015)}]{catalysis15}%
	\BibitemOpen
	\bibfield  {author} {\bibinfo {author} {\bibfnamefont {X.-x.}\ \bibnamefont
			{Xu}},\ }\href {\doibase 10.1103/PhysRevA.92.012318} {\bibfield  {journal}
		{\bibinfo  {journal} {Phys. Rev. A}\ }\textbf {\bibinfo {volume} {92}},\
		\bibinfo {pages} {012318} (\bibinfo {year} {2015})}\BibitemShut {NoStop}%
	\bibitem [{\citenamefont {Hu}\ \emph {et~al.}(2017)\citenamefont {Hu},
		\citenamefont {Liao},\ and\ \citenamefont {Zubairy}}]{catalysis17}%
	\BibitemOpen
	\bibfield  {author} {\bibinfo {author} {\bibfnamefont {L.}~\bibnamefont
			{Hu}}, \bibinfo {author} {\bibfnamefont {Z.}~\bibnamefont {Liao}}, \ and\
		\bibinfo {author} {\bibfnamefont {M.~S.}\ \bibnamefont {Zubairy}},\ }\href
	{\doibase 10.1103/PhysRevA.95.012310} {\bibfield  {journal} {\bibinfo
			{journal} {Phys. Rev. A}\ }\textbf {\bibinfo {volume} {95}},\ \bibinfo
		{pages} {012310} (\bibinfo {year} {2017})}\BibitemShut {NoStop}%
	\bibitem [{\citenamefont {Wang}\ \emph {et~al.}(2015)\citenamefont {Wang},
		\citenamefont {Hou}, \citenamefont {Chen},\ and\ \citenamefont
		{Xu}}]{wang2015}%
	\BibitemOpen
	\bibfield  {author} {\bibinfo {author} {\bibfnamefont {S.}~\bibnamefont
			{Wang}}, \bibinfo {author} {\bibfnamefont {L.-L.}\ \bibnamefont {Hou}},
		\bibinfo {author} {\bibfnamefont {X.-F.}\ \bibnamefont {Chen}}, \ and\
		\bibinfo {author} {\bibfnamefont {X.-F.}\ \bibnamefont {Xu}},\ }\href
	{\doibase 10.1103/PhysRevA.91.063832} {\bibfield  {journal} {\bibinfo
			{journal} {Phys. Rev. A}\ }\textbf {\bibinfo {volume} {91}},\ \bibinfo
		{pages} {063832} (\bibinfo {year} {2015})}\BibitemShut {NoStop}%
	\bibitem [{\citenamefont {Huang}\ \emph {et~al.}(2013)\citenamefont {Huang},
		\citenamefont {He}, \citenamefont {Fang},\ and\ \citenamefont
		{Zeng}}]{qkd-pra-2013}%
	\BibitemOpen
	\bibfield  {author} {\bibinfo {author} {\bibfnamefont {P.}~\bibnamefont
			{Huang}}, \bibinfo {author} {\bibfnamefont {G.}~\bibnamefont {He}}, \bibinfo
		{author} {\bibfnamefont {J.}~\bibnamefont {Fang}}, \ and\ \bibinfo {author}
		{\bibfnamefont {G.}~\bibnamefont {Zeng}},\ }\href {\doibase
		10.1103/PhysRevA.87.012317} {\bibfield  {journal} {\bibinfo  {journal} {Phys.
				Rev. A}\ }\textbf {\bibinfo {volume} {87}},\ \bibinfo {pages} {012317}
		(\bibinfo {year} {2013})}\BibitemShut {NoStop}%
	\bibitem [{\citenamefont {Ma}\ \emph {et~al.}(2018)\citenamefont {Ma},
		\citenamefont {Huang}, \citenamefont {Bai}, \citenamefont {Wang},
		\citenamefont {Bao},\ and\ \citenamefont {Zeng}}]{qkd-pra-2018}%
	\BibitemOpen
	\bibfield  {author} {\bibinfo {author} {\bibfnamefont {H.-X.}\ \bibnamefont
			{Ma}}, \bibinfo {author} {\bibfnamefont {P.}~\bibnamefont {Huang}}, \bibinfo
		{author} {\bibfnamefont {D.-Y.}\ \bibnamefont {Bai}}, \bibinfo {author}
		{\bibfnamefont {S.-Y.}\ \bibnamefont {Wang}}, \bibinfo {author}
		{\bibfnamefont {W.-S.}\ \bibnamefont {Bao}}, \ and\ \bibinfo {author}
		{\bibfnamefont {G.-H.}\ \bibnamefont {Zeng}},\ }\href {\doibase
		10.1103/PhysRevA.97.042329} {\bibfield  {journal} {\bibinfo  {journal} {Phys.
				Rev. A}\ }\textbf {\bibinfo {volume} {97}},\ \bibinfo {pages} {042329}
		(\bibinfo {year} {2018})}\BibitemShut {NoStop}%
	\bibitem [{\citenamefont {Guo}\ \emph {et~al.}(2019)\citenamefont {Guo},
		\citenamefont {Ye}, \citenamefont {Zhong},\ and\ \citenamefont
		{Liao}}]{qkd-pra-2019}%
	\BibitemOpen
	\bibfield  {author} {\bibinfo {author} {\bibfnamefont {Y.}~\bibnamefont
			{Guo}}, \bibinfo {author} {\bibfnamefont {W.}~\bibnamefont {Ye}}, \bibinfo
		{author} {\bibfnamefont {H.}~\bibnamefont {Zhong}}, \ and\ \bibinfo {author}
		{\bibfnamefont {Q.}~\bibnamefont {Liao}},\ }\href {\doibase
		10.1103/PhysRevA.99.032327} {\bibfield  {journal} {\bibinfo  {journal} {Phys.
				Rev. A}\ }\textbf {\bibinfo {volume} {99}},\ \bibinfo {pages} {032327}
		(\bibinfo {year} {2019})}\BibitemShut {NoStop}%
	\bibitem [{\citenamefont {Ye}\ \emph {et~al.}(2019)\citenamefont {Ye},
		\citenamefont {Zhong}, \citenamefont {Liao}, \citenamefont {Huang},
		\citenamefont {Hu},\ and\ \citenamefont {Guo}}]{qk2019}%
	\BibitemOpen
	\bibfield  {author} {\bibinfo {author} {\bibfnamefont {W.}~\bibnamefont
			{Ye}}, \bibinfo {author} {\bibfnamefont {H.}~\bibnamefont {Zhong}}, \bibinfo
		{author} {\bibfnamefont {Q.}~\bibnamefont {Liao}}, \bibinfo {author}
		{\bibfnamefont {D.}~\bibnamefont {Huang}}, \bibinfo {author} {\bibfnamefont
			{L.}~\bibnamefont {Hu}}, \ and\ \bibinfo {author} {\bibfnamefont
			{Y.}~\bibnamefont {Guo}},\ }\href {\doibase 10.1364/OE.27.017186} {\bibfield
		{journal} {\bibinfo  {journal} {Opt. Express}\ }\textbf {\bibinfo {volume}
			{27}},\ \bibinfo {pages} {17186} (\bibinfo {year} {2019})}\BibitemShut
	{NoStop}%
	\bibitem [{\citenamefont {Kumar}\ \emph {et~al.}(2019)\citenamefont {Kumar},
		\citenamefont {Singh}, \citenamefont {Bose},\ and\ \citenamefont
		{Arvind}}]{chandan-pra-2019}%
	\BibitemOpen
	\bibfield  {author} {\bibinfo {author} {\bibfnamefont {C.}~\bibnamefont
			{Kumar}}, \bibinfo {author} {\bibfnamefont {J.}~\bibnamefont {Singh}},
		\bibinfo {author} {\bibfnamefont {S.}~\bibnamefont {Bose}}, \ and\ \bibinfo
		{author} {\bibnamefont {Arvind}},\ }\href {\doibase
		10.1103/PhysRevA.100.052329} {\bibfield  {journal} {\bibinfo  {journal}
			{Phys. Rev. A}\ }\textbf {\bibinfo {volume} {100}},\ \bibinfo {pages}
		{052329} (\bibinfo {year} {2019})}\BibitemShut {NoStop}%
	\bibitem [{\citenamefont {Hu}\ \emph {et~al.}(2020)\citenamefont {Hu},
		\citenamefont {Al-amri}, \citenamefont {Liao},\ and\ \citenamefont
		{Zubairy}}]{zubairy-pra-2020}%
	\BibitemOpen
	\bibfield  {author} {\bibinfo {author} {\bibfnamefont {L.}~\bibnamefont
			{Hu}}, \bibinfo {author} {\bibfnamefont {M.}~\bibnamefont {Al-amri}},
		\bibinfo {author} {\bibfnamefont {Z.}~\bibnamefont {Liao}}, \ and\ \bibinfo
		{author} {\bibfnamefont {M.~S.}\ \bibnamefont {Zubairy}},\ }\href {\doibase
		10.1103/PhysRevA.102.012608} {\bibfield  {journal} {\bibinfo  {journal}
			{Phys. Rev. A}\ }\textbf {\bibinfo {volume} {102}},\ \bibinfo {pages}
		{012608} (\bibinfo {year} {2020})}\BibitemShut {NoStop}%
	\bibitem [{\citenamefont {Tan}\ \emph {et~al.}(2008)\citenamefont {Tan},
		\citenamefont {Erkmen}, \citenamefont {Giovannetti}, \citenamefont {Guha},
		\citenamefont {Lloyd}, \citenamefont {Maccone}, \citenamefont {Pirandola},\
		and\ \citenamefont {Shapiro}}]{ill2008}%
	\BibitemOpen
	\bibfield  {author} {\bibinfo {author} {\bibfnamefont {S.-H.}\ \bibnamefont
			{Tan}}, \bibinfo {author} {\bibfnamefont {B.~I.}\ \bibnamefont {Erkmen}},
		\bibinfo {author} {\bibfnamefont {V.}~\bibnamefont {Giovannetti}}, \bibinfo
		{author} {\bibfnamefont {S.}~\bibnamefont {Guha}}, \bibinfo {author}
		{\bibfnamefont {S.}~\bibnamefont {Lloyd}}, \bibinfo {author} {\bibfnamefont
			{L.}~\bibnamefont {Maccone}}, \bibinfo {author} {\bibfnamefont
			{S.}~\bibnamefont {Pirandola}}, \ and\ \bibinfo {author} {\bibfnamefont
			{J.~H.}\ \bibnamefont {Shapiro}},\ }\href {\doibase
		10.1103/PhysRevLett.101.253601} {\bibfield  {journal} {\bibinfo  {journal}
			{Phys. Rev. Lett.}\ }\textbf {\bibinfo {volume} {101}},\ \bibinfo {pages}
		{253601} (\bibinfo {year} {2008})}\BibitemShut {NoStop}%
	\bibitem [{\citenamefont {Lopaeva}\ \emph {et~al.}(2013)\citenamefont
		{Lopaeva}, \citenamefont {Ruo~Berchera}, \citenamefont {Degiovanni},
		\citenamefont {Olivares}, \citenamefont {Brida},\ and\ \citenamefont
		{Genovese}}]{ill2013}%
	\BibitemOpen
	\bibfield  {author} {\bibinfo {author} {\bibfnamefont {E.~D.}\ \bibnamefont
			{Lopaeva}}, \bibinfo {author} {\bibfnamefont {I.}~\bibnamefont
			{Ruo~Berchera}}, \bibinfo {author} {\bibfnamefont {I.~P.}\ \bibnamefont
			{Degiovanni}}, \bibinfo {author} {\bibfnamefont {S.}~\bibnamefont
			{Olivares}}, \bibinfo {author} {\bibfnamefont {G.}~\bibnamefont {Brida}}, \
		and\ \bibinfo {author} {\bibfnamefont {M.}~\bibnamefont {Genovese}},\ }\href
	{\doibase 10.1103/PhysRevLett.110.153603} {\bibfield  {journal} {\bibinfo
			{journal} {Phys. Rev. Lett.}\ }\textbf {\bibinfo {volume} {110}},\ \bibinfo
		{pages} {153603} (\bibinfo {year} {2013})}\BibitemShut {NoStop}%
	\bibitem [{\citenamefont {Zhang}\ and\ \citenamefont
		{Zhang}(2018)}]{nla-pra-2018}%
	\BibitemOpen
	\bibfield  {author} {\bibinfo {author} {\bibfnamefont {S.}~\bibnamefont
			{Zhang}}\ and\ \bibinfo {author} {\bibfnamefont {X.}~\bibnamefont {Zhang}},\
	}\href {\doibase 10.1103/PhysRevA.97.043830} {\bibfield  {journal} {\bibinfo
			{journal} {Phys. Rev. A}\ }\textbf {\bibinfo {volume} {97}},\ \bibinfo
		{pages} {043830} (\bibinfo {year} {2018})}\BibitemShut {NoStop}%
	\bibitem [{\citenamefont {Birrittella}\ \emph {et~al.}(2012)\citenamefont
		{Birrittella}, \citenamefont {Mimih},\ and\ \citenamefont
		{Gerry}}]{gerryc-pra-2012}%
	\BibitemOpen
	\bibfield  {author} {\bibinfo {author} {\bibfnamefont {R.}~\bibnamefont
			{Birrittella}}, \bibinfo {author} {\bibfnamefont {J.}~\bibnamefont {Mimih}},
		\ and\ \bibinfo {author} {\bibfnamefont {C.~C.}\ \bibnamefont {Gerry}},\
	}\href {\doibase 10.1103/PhysRevA.86.063828} {\bibfield  {journal} {\bibinfo
			{journal} {Phys. Rev. A}\ }\textbf {\bibinfo {volume} {86}},\ \bibinfo
		{pages} {063828} (\bibinfo {year} {2012})}\BibitemShut {NoStop}%
	\bibitem [{\citenamefont {Carranza}\ and\ \citenamefont
		{Gerry}(2012)}]{josab-2012}%
	\BibitemOpen
	\bibfield  {author} {\bibinfo {author} {\bibfnamefont {R.}~\bibnamefont
			{Carranza}}\ and\ \bibinfo {author} {\bibfnamefont {C.~C.}\ \bibnamefont
			{Gerry}},\ }\href {\doibase 10.1364/JOSAB.29.002581} {\bibfield  {journal}
		{\bibinfo  {journal} {J. Opt. Soc. Am. B}\ }\textbf {\bibinfo {volume}
			{29}},\ \bibinfo {pages} {2581} (\bibinfo {year} {2012})}\BibitemShut
	{NoStop}%
	\bibitem [{\citenamefont {Braun}\ \emph {et~al.}(2014)\citenamefont {Braun},
		\citenamefont {Jian}, \citenamefont {Pinel},\ and\ \citenamefont
		{Treps}}]{braun-pra-2014}%
	\BibitemOpen
	\bibfield  {author} {\bibinfo {author} {\bibfnamefont {D.}~\bibnamefont
			{Braun}}, \bibinfo {author} {\bibfnamefont {P.}~\bibnamefont {Jian}},
		\bibinfo {author} {\bibfnamefont {O.}~\bibnamefont {Pinel}}, \ and\ \bibinfo
		{author} {\bibfnamefont {N.}~\bibnamefont {Treps}},\ }\href {\doibase
		10.1103/PhysRevA.90.013821} {\bibfield  {journal} {\bibinfo  {journal} {Phys.
				Rev. A}\ }\textbf {\bibinfo {volume} {90}},\ \bibinfo {pages} {013821}
		(\bibinfo {year} {2014})}\BibitemShut {NoStop}%
	\bibitem [{\citenamefont {Ouyang}\ \emph {et~al.}(2016)\citenamefont {Ouyang},
		\citenamefont {Wang},\ and\ \citenamefont {Zhang}}]{josab-2016}%
	\BibitemOpen
	\bibfield  {author} {\bibinfo {author} {\bibfnamefont {Y.}~\bibnamefont
			{Ouyang}}, \bibinfo {author} {\bibfnamefont {S.}~\bibnamefont {Wang}}, \ and\
		\bibinfo {author} {\bibfnamefont {L.}~\bibnamefont {Zhang}},\ }\href
	{\doibase 10.1364/JOSAB.33.001373} {\bibfield  {journal} {\bibinfo  {journal}
			{J. Opt. Soc. Am. B}\ }\textbf {\bibinfo {volume} {33}},\ \bibinfo {pages}
		{1373} (\bibinfo {year} {2016})}\BibitemShut {NoStop}%
	\bibitem [{\citenamefont {Zhang}\ \emph {et~al.}(2021)\citenamefont {Zhang},
		\citenamefont {Ye}, \citenamefont {Wei}, \citenamefont {Xia}, \citenamefont
		{Chang}, \citenamefont {Liao},\ and\ \citenamefont
		{Hu}}]{pra-catalysis-2021}%
	\BibitemOpen
	\bibfield  {author} {\bibinfo {author} {\bibfnamefont {H.}~\bibnamefont
			{Zhang}}, \bibinfo {author} {\bibfnamefont {W.}~\bibnamefont {Ye}}, \bibinfo
		{author} {\bibfnamefont {C.}~\bibnamefont {Wei}}, \bibinfo {author}
		{\bibfnamefont {Y.}~\bibnamefont {Xia}}, \bibinfo {author} {\bibfnamefont
			{S.}~\bibnamefont {Chang}}, \bibinfo {author} {\bibfnamefont
			{Z.}~\bibnamefont {Liao}}, \ and\ \bibinfo {author} {\bibfnamefont
			{L.}~\bibnamefont {Hu}},\ }\href {\doibase 10.1103/PhysRevA.103.013705}
	{\bibfield  {journal} {\bibinfo  {journal} {Phys. Rev. A}\ }\textbf {\bibinfo
			{volume} {103}},\ \bibinfo {pages} {013705} (\bibinfo {year}
		{2021})}\BibitemShut {NoStop}%
	\bibitem [{\citenamefont {Bartley}\ and\ \citenamefont
		{Walmsley}(2015)}]{njp-2015}%
	\BibitemOpen
	\bibfield  {author} {\bibinfo {author} {\bibfnamefont {T.~J.}\ \bibnamefont
			{Bartley}}\ and\ \bibinfo {author} {\bibfnamefont {I.~A.}\ \bibnamefont
			{Walmsley}},\ }\href {\doibase 10.1088/1367-2630/17/2/023038} {\bibfield
		{journal} {\bibinfo  {journal} {New Journal of Physics}\ }\textbf {\bibinfo
			{volume} {17}},\ \bibinfo {pages} {023038} (\bibinfo {year}
		{2015})}\BibitemShut {NoStop}%
	\bibitem [{\citenamefont {Arvind}\ \emph {et~al.}(1995)\citenamefont {Arvind},
		\citenamefont {Dutta}, \citenamefont {Mukunda},\ and\ \citenamefont
		{Simon}}]{arvind1995}%
	\BibitemOpen
	\bibfield  {author} {\bibinfo {author} {\bibnamefont {Arvind}}, \bibinfo
		{author} {\bibfnamefont {B.}~\bibnamefont {Dutta}}, \bibinfo {author}
		{\bibfnamefont {N.}~\bibnamefont {Mukunda}}, \ and\ \bibinfo {author}
		{\bibfnamefont {R.}~\bibnamefont {Simon}},\ }\href {\doibase
		10.1007/BF02848172} {\bibfield  {journal} {\bibinfo  {journal} {Pramana}\
		}\textbf {\bibinfo {volume} {45}},\ \bibinfo {pages} {471} (\bibinfo {year}
		{1995})}\BibitemShut {NoStop}%
	\bibitem [{\citenamefont {Braunstein}\ and\ \citenamefont {van
			Loock}(2005)}]{Braunstein}%
	\BibitemOpen
	\bibfield  {author} {\bibinfo {author} {\bibfnamefont {S.~L.}\ \bibnamefont
			{Braunstein}}\ and\ \bibinfo {author} {\bibfnamefont {P.}~\bibnamefont {van
				Loock}},\ }\href {\doibase 10.1103/RevModPhys.77.513} {\bibfield  {journal}
		{\bibinfo  {journal} {Rev. Mod. Phys.}\ }\textbf {\bibinfo {volume} {77}},\
		\bibinfo {pages} {513} (\bibinfo {year} {2005})}\BibitemShut {NoStop}%
	\bibitem [{\citenamefont {Adesso}\ and\ \citenamefont
		{Illuminati}(2007)}]{adesso-2007}%
	\BibitemOpen
	\bibfield  {author} {\bibinfo {author} {\bibfnamefont {G.}~\bibnamefont
			{Adesso}}\ and\ \bibinfo {author} {\bibfnamefont {F.}~\bibnamefont
			{Illuminati}},\ }\href {\doibase 10.1088/1751-8113/40/28/S01} {\bibfield
		{journal} {\bibinfo  {journal} {J. Phys. A}\ }\textbf {\bibinfo {volume}
			{40}},\ \bibinfo {pages} {7821} (\bibinfo {year} {2007})}\BibitemShut
	{NoStop}%
	\bibitem [{\citenamefont {Weedbrook}\ \emph {et~al.}(2012)\citenamefont
		{Weedbrook}, \citenamefont {Pirandola}, \citenamefont {Garc\'{\i}a-Patr\'on},
		\citenamefont {Cerf}, \citenamefont {Ralph}, \citenamefont {Shapiro},\ and\
		\citenamefont {Lloyd}}]{weedbrook-rmp-2012}%
	\BibitemOpen
	\bibfield  {author} {\bibinfo {author} {\bibfnamefont {C.}~\bibnamefont
			{Weedbrook}}, \bibinfo {author} {\bibfnamefont {S.}~\bibnamefont
			{Pirandola}}, \bibinfo {author} {\bibfnamefont {R.}~\bibnamefont
			{Garc\'{\i}a-Patr\'on}}, \bibinfo {author} {\bibfnamefont {N.~J.}\
			\bibnamefont {Cerf}}, \bibinfo {author} {\bibfnamefont {T.~C.}\ \bibnamefont
			{Ralph}}, \bibinfo {author} {\bibfnamefont {J.~H.}\ \bibnamefont {Shapiro}},
		\ and\ \bibinfo {author} {\bibfnamefont {S.}~\bibnamefont {Lloyd}},\ }\href
	{\doibase 10.1103/RevModPhys.84.621} {\bibfield  {journal} {\bibinfo
			{journal} {Rev. Mod. Phys.}\ }\textbf {\bibinfo {volume} {84}},\ \bibinfo
		{pages} {621} (\bibinfo {year} {2012})}\BibitemShut {NoStop}%
	\bibitem [{\citenamefont {Adesso}\ \emph {et~al.}(2014)\citenamefont {Adesso},
		\citenamefont {Ragy},\ and\ \citenamefont {Lee}}]{adesso-2014}%
	\BibitemOpen
	\bibfield  {author} {\bibinfo {author} {\bibfnamefont {G.}~\bibnamefont
			{Adesso}}, \bibinfo {author} {\bibfnamefont {S.}~\bibnamefont {Ragy}}, \ and\
		\bibinfo {author} {\bibfnamefont {A.~R.}\ \bibnamefont {Lee}},\ }\href
	{\doibase 10.1142/S1230161214400010} {\bibfield  {journal} {\bibinfo
			{journal} {Open Syst. Inf. Dyn.}\ }\textbf {\bibinfo {volume} {21}},\
		\bibinfo {pages} {1440001, 47} (\bibinfo {year} {2014})}\BibitemShut
	{NoStop}%
	\bibitem [{\citenamefont {Royer}(1977)}]{parity-1977}%
	\BibitemOpen
	\bibfield  {author} {\bibinfo {author} {\bibfnamefont {A.}~\bibnamefont
			{Royer}},\ }\href {\doibase 10.1103/PhysRevA.15.449} {\bibfield  {journal}
		{\bibinfo  {journal} {Phys. Rev. A}\ }\textbf {\bibinfo {volume} {15}},\
		\bibinfo {pages} {449} (\bibinfo {year} {1977})}\BibitemShut {NoStop}%
	\bibitem [{\citenamefont {Yurke}\ \emph {et~al.}(1986)\citenamefont {Yurke},
		\citenamefont {McCall},\ and\ \citenamefont {Klauder}}]{yurke-1986}%
	\BibitemOpen
	\bibfield  {author} {\bibinfo {author} {\bibfnamefont {B.}~\bibnamefont
			{Yurke}}, \bibinfo {author} {\bibfnamefont {S.~L.}\ \bibnamefont {McCall}}, \
		and\ \bibinfo {author} {\bibfnamefont {J.~R.}\ \bibnamefont {Klauder}},\
	}\href {\doibase 10.1103/PhysRevA.33.4033} {\bibfield  {journal} {\bibinfo
			{journal} {Phys. Rev. A}\ }\textbf {\bibinfo {volume} {33}},\ \bibinfo
		{pages} {4033} (\bibinfo {year} {1986})}\BibitemShut {NoStop}%
	\bibitem [{\citenamefont {Braunstein}\ and\ \citenamefont
		{Caves}(1994)}]{braunstein-prl-1994}%
	\BibitemOpen
	\bibfield  {author} {\bibinfo {author} {\bibfnamefont {S.~L.}\ \bibnamefont
			{Braunstein}}\ and\ \bibinfo {author} {\bibfnamefont {C.~M.}\ \bibnamefont
			{Caves}},\ }\href {\doibase 10.1103/PhysRevLett.72.3439} {\bibfield
		{journal} {\bibinfo  {journal} {Phys. Rev. Lett.}\ }\textbf {\bibinfo
			{volume} {72}},\ \bibinfo {pages} {3439} (\bibinfo {year}
		{1994})}\BibitemShut {NoStop}%
	\bibitem [{\citenamefont {Birrittella}\ \emph {et~al.}(2021)\citenamefont
		{Birrittella}, \citenamefont {Alsing},\ and\ \citenamefont
		{Gerry}}]{Birrittella-2021}%
	\BibitemOpen
	\bibfield  {author} {\bibinfo {author} {\bibfnamefont {R.~J.}\ \bibnamefont
			{Birrittella}}, \bibinfo {author} {\bibfnamefont {P.~M.}\ \bibnamefont
			{Alsing}}, \ and\ \bibinfo {author} {\bibfnamefont {C.~C.}\ \bibnamefont
			{Gerry}},\ }\href {\doibase 10.1116/5.0026148} {\bibfield  {journal}
		{\bibinfo  {journal} {AVS Quantum Science}\ }\textbf {\bibinfo {volume}
			{3}},\ \bibinfo {pages} {014701} (\bibinfo {year} {2021})}\BibitemShut
	{NoStop}%
	\bibitem [{\citenamefont {Zavatta}\ \emph {et~al.}(2004)\citenamefont
		{Zavatta}, \citenamefont {Viciani},\ and\ \citenamefont
		{Bellini}}]{Zavatta-2006}%
	\BibitemOpen
	\bibfield  {author} {\bibinfo {author} {\bibfnamefont {A.}~\bibnamefont
			{Zavatta}}, \bibinfo {author} {\bibfnamefont {S.}~\bibnamefont {Viciani}}, \
		and\ \bibinfo {author} {\bibfnamefont {M.}~\bibnamefont {Bellini}},\ }\href
	{\doibase 10.1126/science.1103190} {\bibfield  {journal} {\bibinfo  {journal}
			{Science}\ }\textbf {\bibinfo {volume} {306}},\ \bibinfo {pages} {660}
		(\bibinfo {year} {2004})}\BibitemShut {NoStop}%
	\bibitem [{\citenamefont {Zavatta}\ \emph {et~al.}(2007)\citenamefont
		{Zavatta}, \citenamefont {Parigi},\ and\ \citenamefont
		{Bellini}}]{Zavatta-pra-2007}%
	\BibitemOpen
	\bibfield  {author} {\bibinfo {author} {\bibfnamefont {A.}~\bibnamefont
			{Zavatta}}, \bibinfo {author} {\bibfnamefont {V.}~\bibnamefont {Parigi}}, \
		and\ \bibinfo {author} {\bibfnamefont {M.}~\bibnamefont {Bellini}},\ }\href
	{\doibase 10.1103/PhysRevA.75.052106} {\bibfield  {journal} {\bibinfo
			{journal} {Phys. Rev. A}\ }\textbf {\bibinfo {volume} {75}},\ \bibinfo
		{pages} {052106} (\bibinfo {year} {2007})}\BibitemShut {NoStop}%
\end{thebibliography}
%

\end{document}